\documentclass[manuscript]{aastex}

\usepackage{natbib}
\usepackage{amsmath}
\usepackage{textcomp}
\usepackage[colorlinks=true,citecolor=blue]{hyperref}
\usepackage{graphicx,color}
\usepackage{times}
\usepackage{amssymb}

\newcommand{\lum}{erg\,s$^{-1}$}
\newcommand{\fermi}{{\it Fermi}}
\newcommand{\swift}{{\it Swift}}
\newcommand{\nustar}{{\it NuSTAR}}
\newcommand{\phflux}{\mbox{${\rm \, ph \,\, cm^{-2} \, s^{-1}}$}}
\newcommand{\ergflux}{\mbox{${\rm \, erg \,\, cm^{-2} \, s^{-1}}$}}

\slugcomment{submitted to ApJ}

\shorttitle{A Hard GeV Flare from 3C 279}
\shortauthors{Paliya et al.}

\begin{document}

\title{A HARD GAMMA-RAY FLARE FROM 3C 279 IN 2013 DECEMBER}

\author{Vaidehi S. Paliya$^{1,\,2}$, Chris Diltz$^{3}$, Markus B{\"o}ttcher$^{4,3}$, C. S. Stalin$^{1}$, and David Buckley$^{5}$} 
\affil{$^1$Indian Institute of Astrophysics, Block II, Koramangala, Bangalore-560034, India}
\affil{$^2$Department of Physics, University of Calicut, Malappuram-673635, India}
\affil{$^3$Astrophysical Institute, Department of Physics and Astronomy, Ohio University, Athens, OH 45701, USA}
\affil{$^4$Centre for Space Research, North-West University, Potchefstroom, 2520, South Africa}
\affil{$^5$South African Astronomical Observatory, PO Box 9, Observatory 7935, Cape Town, South Africa}
\email{vaidehi@iiap.res.in}

\begin{abstract}
The blazar 3C 279 exhibited twin $\gamma$-ray flares of similar intensity in 2013 December and 2014 April. In this work, we present a detailed multi-wavelength analysis of the 2013 December flaring event. Multi-frequency observations reveal the uncorrelated variability patterns with X-ray and optical-UV fluxes peaking after the $\gamma$-ray maximum. The broadband spectral energy distribution (SED) at the peak of the $\gamma$-ray activity shows a rising $\gamma$-ray spectrum but a declining optical-UV flux. This observation along with the detection of uncorrelated variability behavior rules out the one-zone leptonic emission scenario. We, therefore, adopt two independent methodologies to explain the SED: a time dependent lepto-hadronic modeling and a two-zone leptonic radiative modeling approach. In the lepto-hadronic modeling, a distribution of electrons and protons subjected to a randomly orientated magnetic field produces synchrotron radiation. Electron synchrotron is used to explain the IR to UV emission while proton synchrotron emission is used to explain the high energy $\gamma$-ray emission. A combination of both electron synchrotron self Compton emission and proton synchrotron emission is used to explain the X-ray spectral break seen during the later stage of the flare. In the two-zone modeling, we assume a large emission region emitting primarily in IR to X-rays and $\gamma$-rays to come primarily from a fast moving compact emission region. We conclude by noting that within a span of 4 months, 3C 279 has shown the dominance of a variety of radiative processes over each other and this reflects the complexity involved in understanding the physical properties of blazar jets in general.
\end{abstract}


\section{Introduction}\label{sec:intro}
Powerful relativistic jets, aligned close to the line of sight to the observer, are the characteristic signature of a special class of active 
galactic nuclei (AGN) called blazars. They are classified as flat spectrum radio quasars (FSRQs) and BL Lac objects. Both the classes are known to exhibit rapid flux and polarization variations \citep[e.g.,][]{1995ARA&A..33..163W,2005A&A...442...97A}, flat radio spectra ($\alpha_r~< 0.5;~\rm{S_{\nu}}~\propto~\rm{\nu}^{-\alpha}$) and superluminal patterns at radio wavelengths \citep{2005AJ....130.1418J}. They dominate the extragalactic high energy GeV $\gamma$-ray sky, as seen by \fermi-Large Area Telescope \citep[\fermi-LAT;][]{2009ApJ...697.1071A}. In the GeV band, FSRQs are more luminous and possess steeper spectral indices than BL Lac objects \citep{2015arXiv150106054A}. A good fraction of blazars are 
also known to be emitters of very high energy (VHE, E $>$ 100 GeV) $\gamma$-ray emission. In the VHE band, as of now, only five FSRQs are known i.e. 3C 279 \citep{2008Sci...320.1752M}, PKS 1222+216 \citep{2011ApJ...730L...8A}, PKS 1510$-$089 \citep{2013A&A...554A.107H}, S3 0218+357 \citep{2014ATel.6349....1M}, and PKS 1441+25 \citep{2015ATel.7433....1M}. 

The broadband spectral energy distribution (SED) of blazars is characterized by two broad peaks. The low frequency peak lies in the radio to soft X-ray frequency range and the high energy peak in the MeV$-$GeV range. In the framework of leptonic jet models, the low frequency emission from blazars is explained as synchrotron emission from non-thermal electrons in the jet. The high energy radiation is believed to be associated with the inverse-Compton (IC) scattering of low energy synchrotron photons from the jet \citep[synchrotron self Compton or SSC,][]{1981ApJ...243..700K}. Though the synchrotron plus SSC models have successfully explained the SEDs of BL Lac objects, reproducing the high energy window of the SED of powerful FSRQs with SSC resulted in the physical parameters far from the equipartition condition, and instead, IC components with seed photons coming from outside the jet \citep[external Compton or EC,][]{1987ApJ...322..650B} are found to be desirable. Apart from leptonic mechanisms, SED of blazars are also reproduced by hadronic or lepto-hadronic emission models \citep[e.g.,][]{2003APh....18..593M,2013ApJ...768...54B}.

The quasar 3C 279 ($z$ = 0.536) is one of the first blazar found to be $\gamma$-ray emitting by the Energetic Gamma-Ray Experiment Telescope ({\it EGRET}) onboard the Compton Gamma-Ray Observatory \citep[{\it CGRO};][]{1992ApJ...385L...1H}. Also, it is the first FSRQ detected in VHE $\gamma$-rays by the Major Atmospheric Gamma-ray Imaging Cherenkov (MAGIC) telescope \citep{2008Sci...320.1752M}. It is known to vary strongly over the entire electromagnetic spectrum \citep[e.g.,][]{2012ApJ...754..114H,1994ApJ...435L..91M,1998ApJ...497..178W}. A change in the optical polarization associated with the $\gamma$-ray flare in 2009 is also reported \citep{2010Natur.463..919A}. Inconsistent patterns of correlation over various energy bands are seen in 3C 279 \citep[e.g.,][]{2008ApJ...689...79C}. At radio wavelengths, 3C 279 has a compact structure and superluminal patterns with apparent speeds as high as $\sim$20c has been noticed from Very Long Baseline Array Observations \citep{2013AJ....146..120L}. From radio observations, the bulk Lorentz factor and viewing angle of the jet flow are estimated as $\Gamma_{\rm j}$ = 15.5 $\pm$ 2.5 and $\Theta_{\rm j}$ = 2$^{\circ}$.1 $\pm$ 1$^{\circ}$.1 \citep{2004AJ....127.3115J,2005AJ....130.1418J}.

3C 279 was detected in an extremely bright state in 2013 December by \fermi-LAT~\citep{2013ATel.5680....1B}. We denote the period 2013 December 14 to 2014 January 3 (MJD 56,640$-$56,660) as high activity phase.  During this period, not only an extremely bright $\gamma$-ray flare was observed, but also the detected $\gamma$-ray spectrum was extremely hard. In addition to that, a significant X-ray spectral break is also observed, thanks to simultaneous monitoring from \nustar~\citep{2013ApJ...770..103H}, and \swift~\citep{2004ApJ...611.1005G}. In this work, using publicly available data, this $\gamma$-ray outburst is studied in detail. Our observational results are consistent with the findings of \citet{2015ApJ...807...79H}, but we here present a different interpretation. Moreover, the observed hard $\gamma$-ray spectrum has been explained as a consequence of Fermi II order acceleration by \citet{2015ApJ...808L..18A}. In Section~\ref{sec:data_red}, the details of the data reduction procedure are presented and we report the results in Section~\ref{sec:results}. The discussion on the obtained results are presented in Section~\ref{sec:dscsn} and we conclude in Section~\ref{sec:summary}. Throughout, we adopt a $\Lambda$CDM cosmology with the Hubble constant $H_0=71$~km~s$^{-1}$~Mpc$^{-1}$, $\Omega_m = 0.27$, and $\Omega_\Lambda = 0.73$.

\section{Multiwavelength observations and Data Reduction}\label{sec:data_red}
\subsection{{\it Fermi}-Large Area Telescope Observations}\label{subsec:fermi}
The $\gamma$-ray data obtained with LAT were collected from 2013 December 14 to 2014 January 3, i.e., for the period of the outburst. We adopt the standard \fermi-LAT data reduction methodology presented in the online documentation\footnote{http://fermi.gsfc.nasa.gov/ssc/data/analysis/scitools/python\_tutorial.html} and here it is described in brief \citep[see also][]{2015ApJ...803...15P}. In the energy range of 0.1$-$300 GeV, only events belonging to the SOURCE class are selected. To limit contamination from Earth limb $\gamma$-rays, photons arriving from zenith angles $>$ 100$^{\circ}$ are rejected. The collected LAT data are analyzed with the unbinned likelihood method included in the pylikelihood library of the standard ScienceTools package (v9r33p0) along with the use of post-launch instrument response functions P7REP\_SOURCE\_V15. The photons are extracted from a region of interest (ROI) centered on 3C 279 and having a radius of 10$^{\circ}$. The source model consists of 3C 279 and all the point sources from the third \fermi-LAT catalog \citep[3FGL;][]{2015ApJS..218...23A} that fall within 15$^{\circ}$ of the source. The spectral shapes of all the sources are adopted from the 3FGL catalog and the associated parameters, except scaling factor, are left free to vary for the objects lying within the ROI. For the sources lying between 10$^{\circ}$ to 15$^{\circ}$, all the spectral parameters are kept fixed to the 3FGL catalog values. A maximum likelihood (ML) test statistic TS =  2$\Delta \log (\mathcal{L}$) where $\mathcal{L}$ represents the likelihood function between models with and without a point source at the position of the source of interest, is computed to determine the significance of the $\gamma$-ray signal. A first run of the ML analysis is performed over the period of interest and all the sources with TS $<$ 25\footnote{A TS of 25 roughly corresponds to 5$\sigma$ detection \citep{1996ApJ...461..396M}.} are removed. This updated model is then used for light curve and spectral analysis. Though 3C 279 is modeled by a logParabola model in the 3FGL catalog, a power law (PL) model is used to generate light curves, as the PL indices obtained from this model show smaller statistical uncertainties when compared to those obtained from complex model fits. The source is considered to be detected if TS $>$ 9 which corresponds to $\sim$3$\sigma$ detection. We do not consider the bins with TS $<$ 9 and/or $\Delta F_{\gamma}/F_{\gamma} > 0.5$, where $\Delta F_{\gamma}$ is the error estimate in the flux $F_{\gamma}$, in the analysis. The measured fluxes have energy dependent systematic uncertainties of around 10\% below 100 MeV, decreasing linearly in log(E) to 5\% in the range between 316 MeV and 10 GeV and increasing linearly in log(E) up to 15\% at 1 TeV\footnote{http://fermi.gsfc.nasa.gov/ssc/data/analysis/LAT\_caveats.html}. All errors associated with the LAT data analysis are the 1$\sigma$ statistical uncertainties.

\subsection{\nustar~Monitoring}
The hard X-ray focusing telescope \nustar~observed 3C 279 twice for a total elapsed time of $\sim$80 ksec each on 2013 December 16 and 31. The data are filtered and cleaned for background events using the \nustar~Data Analysis Software (NUSTARDAS) version 1.4.1 and  \nustar~calibration files updated on 2014 November 14. The tool {\tt nuproducts} is used to extract light curves and spectra for the two focal plane modules (FPMA and FPMB). To extract the source spectra, a circular 
region of 30$^{\prime\prime}$ centered at 3C 279 is selected, whereas background region is chosen as a circle of 70$^{\prime\prime}$ radius, free from contaminating sources. The spectra  are binned to have atleast 20 counts per bin, to perform spectral fitting. In the energy range of 3$-$79 keV, the light curves are generated by applying 3 ksec binning, summing FPMA and FPMB count rates and subtracting the background.
 
\subsection{{\it Swift} Observations}\label{subsec:swift}
The \swift-XRT data are processed using standard procedures ({\tt xrtpipeline v.0.13.0}) with the XRTDAS software package (v.3.0.0) within the HEASOFT (6.16) and the calibration database updated on 2014 November 12. Only XRT event grades 0$-$12 in the photon counting mode are used. Event files are summed to extract the energy spectrum. Whenever source count rate increases a threshold of 0.5 counts s$^{-1}$, to avoid pile up effect, annular regions centered at the source position are selected to extract the source and the background spectra. The source spectra are extracted from an annular region of inner and outer radii 5$^{\prime\prime}$ and 65$^{\prime\prime}$, respectively, while the background region is chosen as an annular region with inner and outer radii 130$^{\prime\prime}$ and 230$^{\prime\prime}$, respectively \citep[see e.g.,][]{2013ApJS..207...28S}. Exposure maps are combined using the tool {\tt ximage} and we generate the ancillary response files using the task {\tt xrtmkarf}. The task {\tt grppha} is used to bin the source spectra to have at least 20 counts per bin. Spectral fitting is performed using XSPEC \citep{1996ASPC..101...17A} and by adopting an absorbed power law \citep[N$_{\rm H}$ = 2.05 $\times$ 10$^{20}$ cm$^{-2}$;][]{2005A&A...440..775K}. The uncertainties are calculated at 90\% confidence level.

{\it Swift}-UVOT observations are summed using {\tt uvotimsum} and the task {\tt uvotsource} is used to extract the parameters. A circle of 5$^{\prime\prime}$ radius centered at 3C 279 is chosen as source region, while the background events are extracted from a circular region of 1$^{\prime}$ radius free from contaminating sources. Correction for galactic extinction is done by following \citet{2011ApJ...737..103S} and the corrected magnitudes are converted to flux units using the zero points and conversion factors of \citet{2011AIPC.1358..373B}.

\subsection{SMARTS Observations}\label{subsec:smarts}
Small and Moderate Aperture Research Telescope System (SMARTS) at the Cerro Tololo Inter-American Observatory located at Chile has been observing a sample of \fermi-LAT detected AGNs in optical and near-IR (B, V, R, J, and K bands). The details on the data reduction procedure can be found in \citet{2012ApJ...756...13B}. Following \citet{2011ApJ...737..103S}, the data in all the filters are corrected for galactic extinction  and then converted to flux units using the zero points of \citet{1998A&A...333..231B}.

\subsection{Steward Observatory Monitoring}\label{subsec:stewards}
Details of the data reduction and calibration procedures of the photometric and polarimetric observations taken from Steward observatory at the University of Arizona are presented in \citet{2009arXiv0912.3621S}. The data are corrected for galactic extinction following \citet{2011ApJ...737..103S} and the flux conversion is done using the zero points of \citet{1998A&A...333..231B}.
\section{Results}\label{sec:results}
\subsection{Multi-band Temporal Variability}\label{subsec:mw_var}
\begin{figure*}
\hbox{
      \includegraphics[width=\columnwidth]{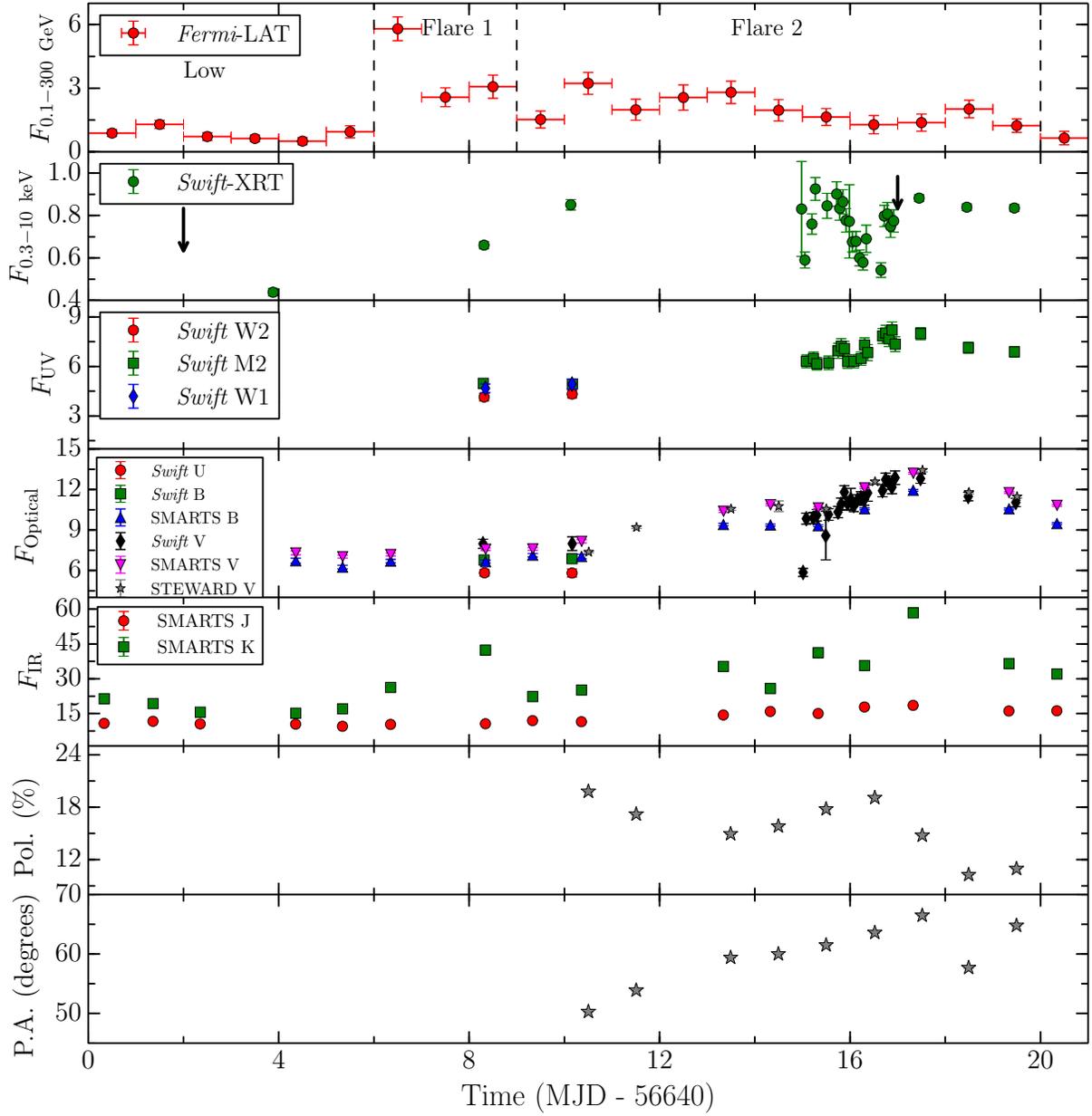}
     }
\caption{Multi-frequency light curves of 3C 279 covering the $\gamma$-ray flaring period. The units of \fermi-LAT and {\it Swift}-XRT data points are 10$^{-6}$ \phflux~and counts s$^{-1}$, respectively. Optical-UV and IR luxes have units of 10$^{-12}$ erg cm$^{-2}$ s$^{-1}$ . Black downward arrows in the second panel from the top coprrespond to the \nustar~monitoring epochs.}\label{fig:mw_lc}
\end{figure*}

The period 2013 December 14 to 2014 January 3 (MJD 56,640$-$56,660) is selected to study the giant $\gamma$-ray outburst of 3C 279 in detail. The multi-frequency light curves covering the data from IR to $\gamma$-rays as well as the optical polarization measurements, are shown in Figure~\ref{fig:mw_lc}. In this plot, LAT data points are one day binned and the observations in other wavelengths correspond to one point per observation. The period of high activity is divided into three sub-periods; Low activity (MJD 56,640$-$56,646), Flare 1 (MJD 56,646$-$56,649), and Flare 2 (MJD 56,649$-$56,660). These sub-periods are selected taking into account the availability of contemporaneous observations in all the energy bands. Lack of the observations during the peak of the $\gamma$-ray flare precludes us to study the nature of the source in the X-ray band, however, a bright X-ray flare is seen during Flare 2 activity phase. Interestingly, available optical-UV observations seem to show a slow continuous rise, irrespective of the $\gamma$-ray flaring activity, and peaks during Flare 2 period. Such uncorrelated variability behavior is difficult to explain on the basis of widely accepted single zone leptonic emission scenario.

\begin{figure*}
\hbox{
      \includegraphics[width=\columnwidth]{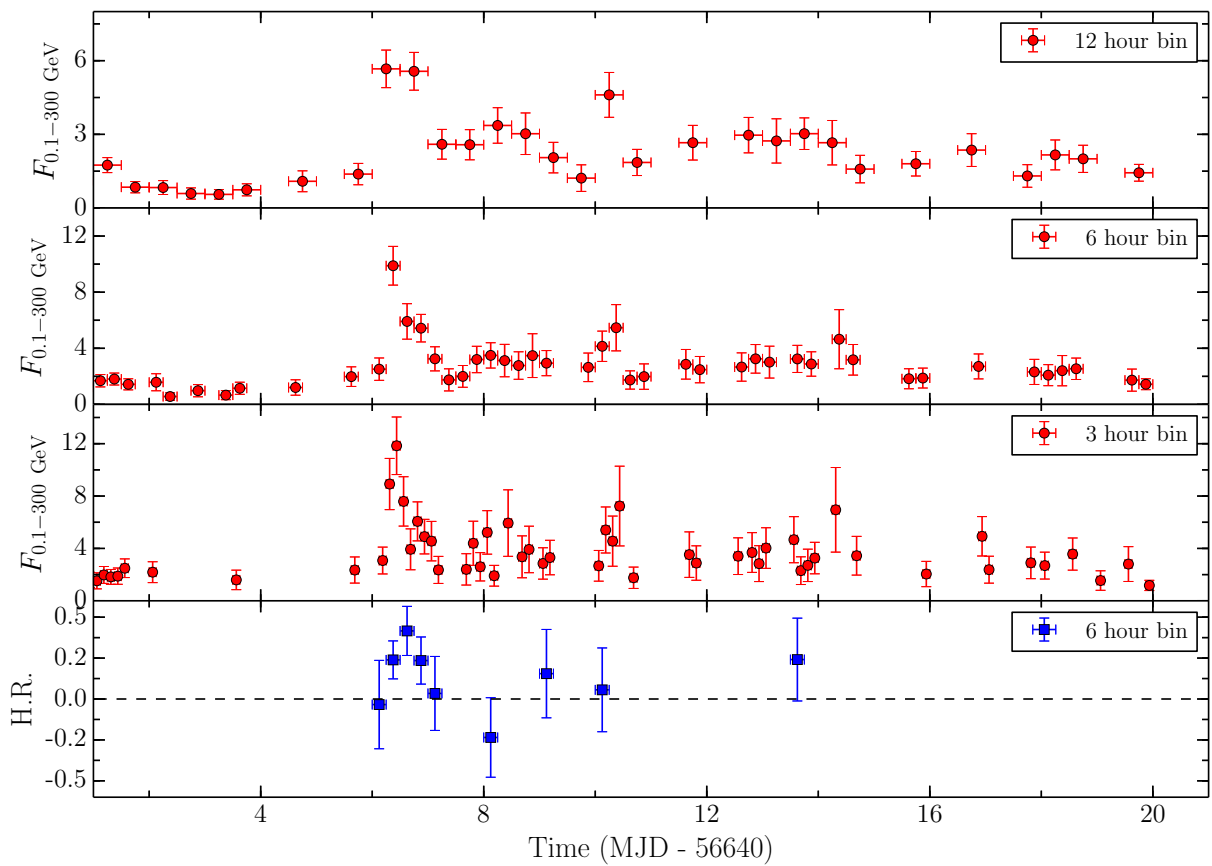}
     }
\caption{Fine time binned \fermi-LAT light curve of 3C 279. The adopted time binning are 12 hr, 6 hr, and 3 hr (top three panels). The flux units are same as in Figure~\ref{fig:mw_lc}. The bottom panel represents variation of the hardness ratio (see equation~\ref{eq:HR}) for 6 hr binned data, as a function of time.}\label{fig:rapid}
\end{figure*}

Availability of good $\gamma$-ray photon statistics has allowed us to search for short time scale variability by using finer time bins. For this, 12 hr, 6 hr, and 3 hr binned $\gamma$-ray light curves are generated covering the period of high activity. They are shown in Figure~\ref{fig:rapid}. These light curves are searched for short time variability using the following equation
\begin{equation}\label{eq:flux-double}
F(t) = F(t_0).2^{(t-t_0)/\tau}
\end{equation}
where $F(t)$ and $F(t_0)$ are the fluxes at time $t$ and $t_0$, respectively, and $\tau$ is the characteristic doubling/halving time scale. By considering the uncertainties in the flux values, we also apply the condition that the difference in flux at the epochs $t$ and $t_0$ is at least significant at the 3$\sigma$ level. The shortest $\gamma$-ray flux doubling time derived using this method is 3.04 $\pm$ 0.77 hours ($\sim$5$\sigma$ significance). Moreover, the data are also analyzed using the time bins defined as good time intervals (GTI). Using this method, in the energy range of 0.1$-$300 GeV, the highest $\gamma$-ray flux and the associated photon index are found on MJD 56,646.48 (GTI $\sim$13 min) and having a value of (1.22 $\pm$ 0.25) $\times$ 10$^{-5}$ \phflux~and 1.70 $\pm$ 0.13 respectively. The measured flux is comparable to that observed from 3C 279 during 2014 April outburst \citep{2015ApJ...803...15P}, however, less than its recent flaring activity in 2015 June \citep{2015ApJ...808L..48P}. Also, the obtained spectral shape is the hardest ever observed from this source and this suggests that at the peak of the $\gamma$-ray flare, 3C 279 was a probable candidate to detect VHE emission. Moreover, we calculate hardness ratio (HR) which is defined as
\begin{equation}\label{eq:HR}
{\rm HR} = \frac{F_{\rm H} - F_{\rm S}}{F_{\rm H} + F_{\rm S}},
\end{equation}
where $F_{\rm S}$ and $F_{\rm H}$ are 6 hr binned $\gamma$-ray fluxes in 0.1$-$1 GeV and 1$-$300 GeV energy range, respectively. The variation of HR as a function of time is shown in the bottom panel of Figure~\ref{fig:rapid} and as can be seen, the HR has the highest value at the peak of the flare. Furthermore, the number of HR measurements is significantly less than the number of points in 6 hr binned 0.1$-$300 GeV light curve. This is only due to non-detection at E$>$1 GeV at those epochs.

\begin{figure*}
\hbox{
      \includegraphics[width=\columnwidth]{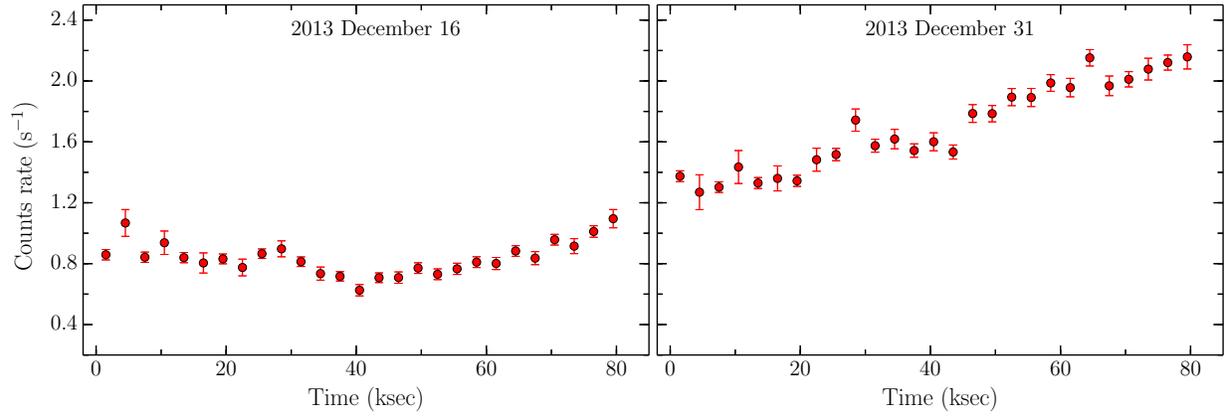}
     }
\caption{Background subtracted 3$-$79 keV light curves of 3C 279, extracted from the \nustar~observations. The FPMA and FPMB count rates are summed and 3 ksec binning is applied.}\label{fig:nustar}
\end{figure*}

The hard X-ray light curves are generated by applying a binning of 3 ksec to the \nustar~data in the energy range of 3$-$79 keV and are shown in Figure~\ref{fig:nustar}. The chi-square probability \citep[e.g.,][]{2010ApJ...722..520A} that the source has shown variations is $>$99\% for both \nustar~observations. The source has also shown significant flux variation between the two epochs.

\swift~has performed 22 observations of 3C 279 on MJD 56,656 and 56,657 (2013 December 30, 31). Such densely sampled observations have revealed not only a bright X-ray flare but also an extremely fast X-ray variability. The shortest X-ray flux doubling time, estimated using Equation~\ref{eq:flux-double}, is 2.89 $\pm$0.67 hours measured on MJD 56,656 with  $\sim$4$\sigma$ confidence. Interestingly, as can be seen in Figure~\ref{fig:mw_lc}, this does not correspond to any $\gamma$-ray flare. This is probably the first report of such a fast X-ray variability seen from 3C 279. In the energy range of 0.3$-$10 keV, the highest X-ray flux is measured on MJD 56,655 and having a value of 3.92$^{+0.49}_{-0.44}~\times$ 10$^{-11}$ \ergflux. The associated photon index is 1.50$^{+0.17}_{-0.16}$. This corresponds to an isotropic X-ray luminosity of $\sim$3.6 $\times$ 10$^{46}$ \lum.

\subsection{Highest Energy Gamma-ray Photon}\label{subsec:highest_gamma_photon}
The energy of the highest energy photon is determined by analyzing the LAT data using event class CLEAN along with the use of the tool {\tt gtsrcprob}. We find the highest energy photon of 26.11 GeV detected on 2013 December 16 (MJD 56,642.26032) at 0$^{\circ}$.018 away from 3C 279, with 99.98\% probability that the event belongs to the source.

\subsection{Gamma-ray Spectral Analysis}\label{subsec:gamma_spec}
\begin{figure*}
\hbox{
      \includegraphics[width=\columnwidth]{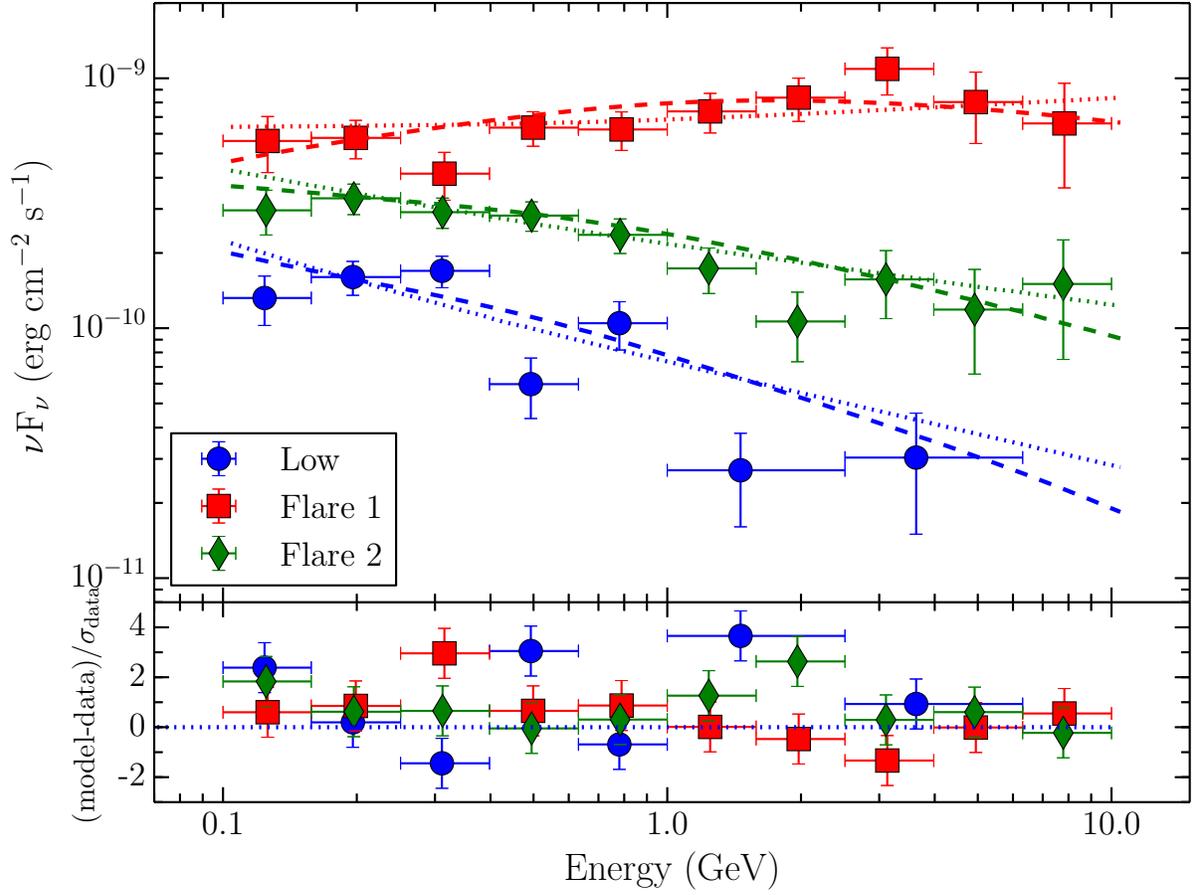}
     }
\caption{Gamma-ray flux distributions of 3C 279 during different activity states. Power law and logParabola models are shown with dotted and dashed lines, respectively. The residuals in the lower panel are with respect to the power law model.}\label{fig:gamma_spec}
\end{figure*}

 We generate the $\gamma$-ray spectra for all three periods, namely Low activity, Flare 1, and Flare 2. Two spectral models: power law ($dN/dE \propto E^{\Gamma_{\gamma}}$), where $\Gamma_{\gamma}$ is the photon index and logParabola ({ $dN/dE \propto (E/E_{\rm o})^{-\alpha-\beta log({\it E/E_{\rm o}})}$}, where $E_{\rm o}$ is an arbitrary reference energy fixed at 300 MeV, $\alpha$ is the photon index at $E_{\rm o}$ and $\beta$ is the curvature index which defines the curvature around the peak) are adopted to analyze the $\gamma$-ray spectral shape. We calculate the test statistic of the curvature $TS_{\rm curve}$ = 2(log $\mathcal{L}$(LogParabola) $-$ log $\mathcal{L}$(power-law)) to test for the presence of curvature. A $TS_{\rm curve}$ $>$16 suggests for the presence of significant curvature \citep{2012ApJS..199...31N}. The fitting parameters are given in Table~\ref{tab:gamma_spec} and the resultant SEDs are shown in Figure~\ref{fig:gamma_spec}. No statistically significant curvature is found. At the peak of the $\gamma$-ray flare, the derived $\gamma$-ray spectral shape is hard and is well explained by a power law model.

\subsection{Spectral Energy Distributions}\label{subsec:sed}
\begin{figure*}
\hbox{
      \includegraphics[width=\columnwidth]{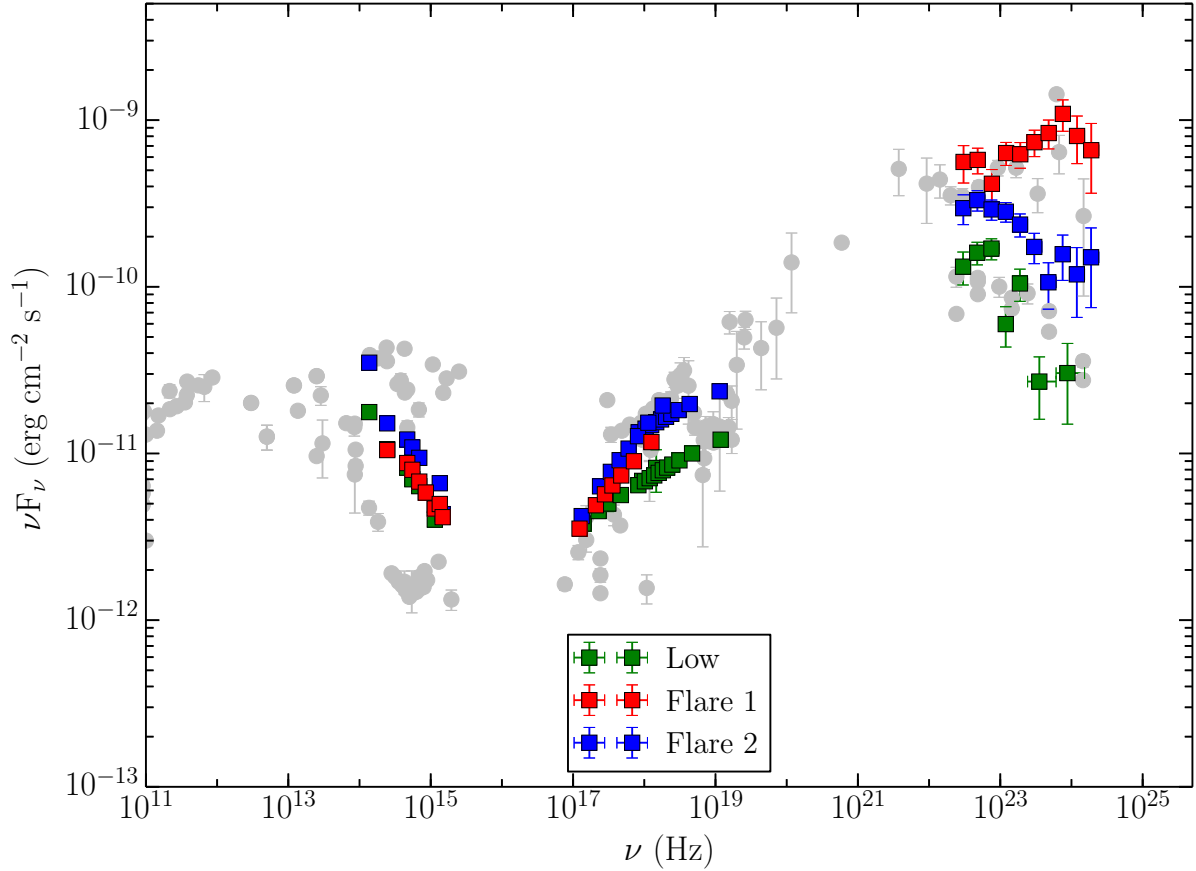}
     }
\caption{\fermi-LAT SEDs of 3C 279 during different activity states. The green data points represent the SED for the low activity state, whereas red and blue points correspond to Flare 1 and Flare 2 periods. Silver gray points represent the archival observations.}\label{fig:sed_all}
\end{figure*}

The broadband SEDs of 3C 279 are generated during three sub-periods and shown in Figure~\ref{fig:sed_all}. We average the flux over each of the three time intervals and the derived values are presented in Table~\ref{tab:sed_flux}, except for \fermi-LAT data which are given in Table~\ref{tab:gamma_spec}. The broadband SEDs are reproduced considering the models presented in Section~\ref{subsubsec:model}.
\subsubsection{Model Setup}\label{subsubsec:model}

We consider a one-zone lepto-hadronic model in which a population of relativistic protons is continuously injected with a power-law distribution $Q_{p} (\gamma_{p}) = Q_{0} \gamma_{p}^{-q_{p}} H(\gamma_{p}; \gamma_{p, min}, \gamma_{p, max})$ into a spherical emission region of comoving radius $R$ and a randomly oriented magnetic field of strength $B$. Here, $H(x; a, b)$ is the Heaviside function defined as $H = 1$ if $a \le x \le b$ and $H = 0$ otherwise. The size of the emission region is constrained through the variability time scale, $\Delta t_{var}$, using the relation $R \leq c\Delta t_{var}/(1+z)$, where $z$ denotes the redshift to the source. The emission region moves along the jet with relativistic bulk Lorentz factor $\Gamma$, \citep[see][]{Diltz2015}. Following the injection, the electrons and protons emit synchrotron radiation from the radio to high energy $\gamma$-rays. Large magnetic fields, $B \geq 10~G$, are necessary in order for the protons to produce significant synchrotron radiation in the broadband emission of relativistic jets and to ensure that the proton Larmor radius is confined to within the size scale of the emission region, $R \approx 10^{15} \ cm$. The time evolution of the injected particle distributions are modeled through separate Fokker-Planck equations. The proton Fokker-Planck equation incorporates losses due to synchrotron, pion-production and adiabatic processes. With the large magnetic fields necessary for the fits, only synchrotron losses are taken into consideration for the electron/positron Fokker-Planck equation.   
 
With the proton distribution and the photon fields, we compute the pion production rates based on the photo-hadronic interaction cross section between protons and photons. The total proton-photon cross section is divided into different components, corresponding to separate channels through which the neutral and charged pions are produced. These channels include: direct resonances (such as the $\Delta$ resonance), higher resonances, direct single-pion production and multi-pion production. We use analytic expressions for the neutral and charged pion production rates in these different channels in order to determine the overall neutral and charged pion distributions \citep{Hummer10}. The production rates are used as injection terms in a Fokker-Planck equation. Because of the short decay time for the neutral pions, $t_{\rm decay}\approx10^{-17}$s, we assume that they decay instantaneously into photons. The charged pions emit synchrotron radiation before they decay to produce charged muons and neutrinos. The muons then follow their own Fokker-Planck equation, emitting synchrotron radiation before decaying into electron/positron pairs and neutrinos. Both Fokker-Planck equations of the secondary particles take into account synchrotron radiative losses, injection rate, escape and decay time. The electrons/positrons production rates resulting from the decay of charged muons serves as an additional injection term for the electron/positron distribution \citep[see][]{Diltz2015}. The coupled Fokker-Planck equations are numerically solved using the Crank-Nichelson scheme to produce equilibrium distributions for each particle species. With these particle distributions, we compute the broadband spectral energy distribution from synchrotron emission of protons, pions, muons and secondary electrons/positrons as well as Compton scattering of electrons/positrons. Since the pion-production rate dominates over the loss rate from Bethe-Heitler pair production, the latter is not expected to play an important role in our model and, therefore, omitted.

The particle distributions can also interact with magnetohydrodynamic waves in the emission region. A resonant interaction between the particle and the transverse component of the electric field of the MHD wave takes place when the Doppler-shifted wave frequency is a constant multiple of the particle gyrofrequency in the particle guiding center frame. The particle will observe either an accelerating or decelerating electric field in the transverse direction of motion over a fraction of the cyclotron period, resulting in an increase or decrease in energy. The particle gyro-resonant interactions with MHD waves causes the particle distributions to diffuse in energy, pushing particles to higher and lower energies. This energy diffusion typically causes the particle distributions to have a pronounced curvature in the energy spectrum. Increased curvature has been reported in the high energy spectral components in blazar modeling \citep[e.g.][]{2010ApJ...710.1271A}. A MHD wave spectral index of $p = 2$ is used to model the particle diffusion in this study.

To explain the flaring states, we assume that the protons are energized as a shock front within the jet becomes compressed. The increasing shock strength leads to more efficient proton acceleration and, thus, a harder proton spectrum. The harder proton spectrum makes the synchrotron spectrum harder, which explains the harder synchrotron spectrum in the \swift-XRT and \nustar~measurements, observed during the flare state. As the protons cross the shock front, they encounter increased turbulence down stream. The protons interact with this increased turbulence downstream which thereby increases the efficiency through which the protons gain and lose energy through Fermi II acceleration. The increased efficiency of stochastic acceleration causes the particle distribution to diffuse in energy, changing the spectral curvature of the synchrotron spectrum. The increased acceleration efficiency also affects the pions, muons and electrons/positrons generated from photohadronic interactions. A combination of a harder spectral index and an increase in the stochastic acceleration of the protons causes an increase in muon emission around $\sim 100$ GeV. There is no observed increase in the optical emission for the initial flare. In the context of hadronic modeling, this implies that the electron population responsible for the synchrotron emission seen at optical wavelengths is unaffected by this perturbation.

Following the initial $\gamma$-ray flare, there is a second flare in the optical, X-ray and $\gamma$-ray bandpasses. This implies that both particle populations are affected and contribute to this flare. We model a perturbation to the electron injection spectrum in which the injection luminosity, electron spectral index, and the efficiency of stochastic acceleration change. The proton distribution goes through the exact same perturbation setup as the first flare, but with a smaller amplitude. The protons and electrons cross the shock front, obtain changes in their spectral indices and diffuse in energy from the increased stochastic acceleration. Given the physical scenario outlined, for the first flare, we modify the proton spectral index, injection luminosity and the acceleration efficiency using the following relations:

\begin{equation}
	q_{p} (t) = q_{p, 0} + K_{q} \cdot e^{-(t - t_{0})^{2}/2 \sigma^{2}}
\end{equation}

\begin{equation}
	L_{inj} (t) = \frac{L_{inj, 0}}{1 + K_{L} \cdot e^{-(t - t_{0})^{2}/2 \sigma^{2}}}
\end{equation}

\begin{equation}
	t_{acc} (t) = \frac{t_{acc, 0}}{1 + K_{acc} \cdot e^{-(t - t_{0})^{2}/2 \sigma^{2}}}
\end{equation}

\noindent where the constant $K$ denotes the strength of the perturbation for the particle spectral index, injection luminosity, and acceleration timescale. The terms $q_{p, 0}$, $L_{inj, 0}$, and $t_{acc, 0}$ denote the particle spectral index, injection luminosity and stochastic acceleration timescale during quiescence, respectively. The variable $\sigma$ represents the duration of the perturbation in the comoving frame and $t_{0}$ denotes the time where the perturbation peaks in our simulation. The value of $\sigma$ is related to the propagation time of the shock. For the size scale, $R$, of the emission region yields a sigma value of $\sigma \sim t \approx R/v_{sh}$, where $v_{sh}$ denotes the size scale of the shock. The value of $\sigma$ is assumed to be the same for all three perturbations for a given flare. The values for $\sigma$ are chosen to reproduce the rise and decay times of the light curves in the different bands for both flares. These perturbations ensure that the proton synchrotron produces a harder spectrum and an increased luminosity in order to explain the first $\gamma$ ray flare centered at MJD 56,646. The electron and proton perturbation in the second flare, MJD 56,656, is modeled with the same Gaussian functions in time as the protons for the first perturbation. Negative values for the spectral index perturbation, $K_{q}$ indicate a spectral hardening, while a positive value gives a spectral softening. Conversely, a positive value for the injection luminosity perturbation indicate a drop in the particle luminosity, while a negative value indicates an increase.

An alternate approach to reproduce the 2013 December flare can be the use of two-zone leptonic emission scenario. In this approximation, a large region emits primarily in IR to X-rays with some contribution in the $\gamma$-ray band, whereas a small fast moving emission region emits predominantly in the $\gamma$-ray energy regime. The relatively slow variations seen in the IR to X-ray fluxes during the $\gamma$-ray activity period supports their origin from a large emission region. On the other hand, a fast $\gamma$-ray flare could be emitted by a smaller emitting region. This approach is similar to that adopted by \citet{2011A&A...534A..86T} to explain very fast VHE variations and a hard GeV spectrum observed from FSRQ PKS 1222+216. During the 2013 December flare, the VHE monitoring of 3C 279 from the ground based Cherenkov telescopes was not possible due to full moon period. Therefore, unlike PKS 1222+216, we could not constrain the location of the $\gamma$-ray emitting region. This is due to the fact that a confirmed VHE detection will rule out the possibility of the inside BLR origin of the $\gamma$-rays. Therefore, in this work, we present the results according to both inside and outside BLR location of the fast moving small emission region. The precise measurement of the distance of the emission regions from central engine is not very important, since the radiation fields of both BLR and IR-torus remain uniform in the comoving frame, as long as the emission region is inside the respective components \citep{2009MNRAS.397..985G}. Moreover, we assume that both emission regions do not interact with each other \citep[see e.g.,][]{2011A&A...534A..86T}. The size of the large emission region is constrained by adopting it to cover entire jet cross-section with jet semi opening angle of 0.1 rad. The size of the small emission region is derived from the observed fastest $\gamma$-ray variability. To model the broadband SEDs, we follow the guidelines presented in \citet{2009MNRAS.397..985G} and \citet{2009herb.book.....D} and here we describe it in brief. The emission region moves with a bulk Lorentz factor $\Gamma$ and is assumed to be filled with electrons having smooth broken power law energy distribution
\begin{equation}
 N(\gamma)  \, = \, N_0\, { (\gamma_{\rm brk})^{-p} \over
(\gamma/\gamma_{\rm brk})^{p} + (\gamma/\gamma_{\rm brk})^{q}},
\end{equation}
where $p$ and $q$ are the energy indices before and after the break energy ($\gamma_{\rm brk}$), respectively. The synchrotron and SSC emissions are calculated under the assumption of a uniform but tangled magnetic field \citep[e.g.,][]{2008ApJ...686..181F}. The electrons also scatter the photons entering from the accretion disk, the BLR, and the dusty torus via EC process \citep{2009MNRAS.397..985G,2009herb.book.....D}. The jet powers are derived following the prescriptions of \citet{2008MNRAS.385..283C}. The leptonic model used here, does not consider radiative losses to calculate particle spectrum.

\subsubsection{SED Modeling Results}\label{subsubsec:sed_results}
We perform a parameter study to provide a rough fit to the average SED of 3C 279 by running our time-dependent lepto-hadronic code with time-independent input parameters and waiting for the particle and photon spectra to approach equilibrium. To reproduce the equilibrium solutions quickly, we set the time step initially to $\sim 10^{7}$s. This time step is larger than the radiative and acceleration time scales that determine the evolution of the particle distributions in the Fokker-Planck equations. The implicit Crank-Nichelson scheme, used to numerically solve the Fokker-Planck equations, ensures that the simulation converges to a stable solution after a few time steps.

Given the large set of input parameters, we try to constrain as many input parameters as possible from observational data. We have several parameters that we can constrain by observations. The redshift is given as $z = 0.536$. With the superluminal motion speed, we set a lower limit to the bulk Lorentz factor, $\Gamma > 18$. The observing angle is set by using the relation $\theta_{\rm obs} = 1/\Gamma$ so that $\delta = \Gamma$. The variability timescale for the $\gamma$-ray flare is given by $\Delta t_{\rm var, obs}\sim1.5$ hr. The luminosities of the accretion disk and the broad line region (BLR) are $L_{\rm disk} = 2.0 \times 10^{45}$~erg~s$^{-1}$ \citep{1999ApJ...521..112P}, and $L_{\rm BLR} = 2.0 \times 10^{44}$~erg~s$^{-1}$ (assuming BLR to reprocess 10\% of the disk luminosity). From the variability time scale, we can constrain the location of the emission region along the jet, $R_{\rm axis} \sim 2 \, \Gamma^{2} \, c \, t_{v} / (1 + z) \approx 5.84 \times 10^{16}$ cm. With the luminosity of the BLR, we can determine the characteristic size of the BLR using the luminosity-radius relation \citep{Bentz13}. 

For the parameters that are not directly constrained by observations, we perform a "fit-by-eye" method to determine the values of the remaining set. The unconstrained parameters are adjusted until a reasonable fit to the SED is obtained. In the context of lepto-hadronic modeling, the X-ray to soft and intermediate $\gamma$-ray emission is explained by proton synchrotron radiation, while the radio to UV emission is best explained by synchrotron radiation from electrons/positrons. We also require a spectral component due to electron synchrotron self Compton, in order to improve the fits in the X-ray band. For our lepto-hadronic model, we require that the proton distribution and the magnetic field are constrained such that muon and pion synchrotron emission is no longer negligible in the SED fitting \citep[see][]{Diltz2015}. As a result, the high energy emission beyond $10~GeV$ is explained by synchrotron emission from muons. From the fits, we find the magnetic field to be roughly $\sim30$ G. The magnetic field allows the jet to become particle dominated with an equipartition parameter between the particle kinetic luminosity and magnetic field of $\epsilon_{pB} \approx 2.0 \times 10^{3}$. The value selected for $B$ also ensures that the X-ray flux for 3C 279 in the quiescent state is due to a combination of proton synchrotron and electron synchrotron self Compton emission. A full list of parameters for the SED of 3C 279 is given in Table \ref{tab:parameters}.

\begin{figure*}
\hbox{\hspace{-1.5cm}
      \includegraphics[width=10.0cm]{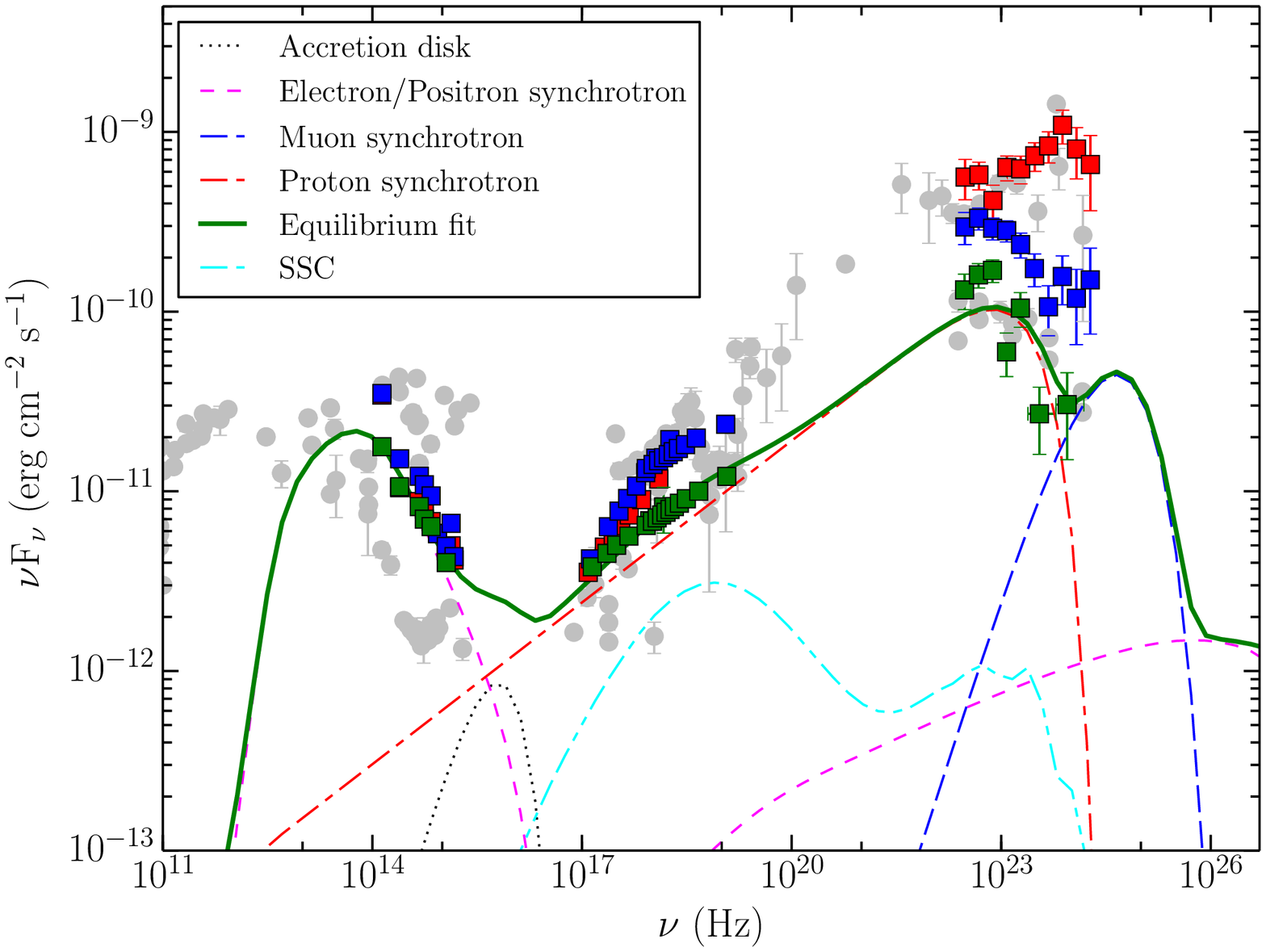}
      \includegraphics[width=10.0cm]{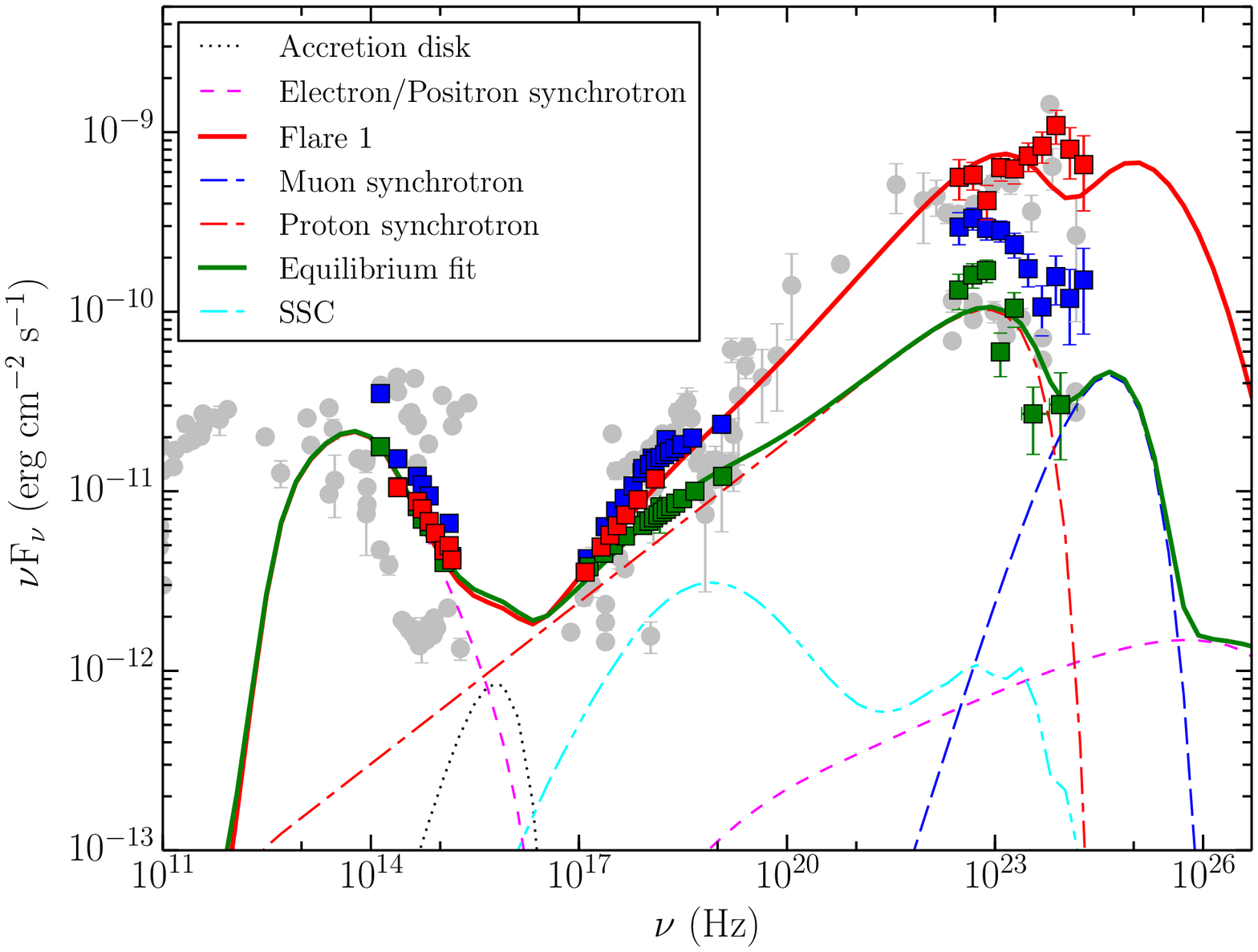}
     }
     \centering
\hbox{\hspace{3.0cm}
      \includegraphics[width=10.0cm]{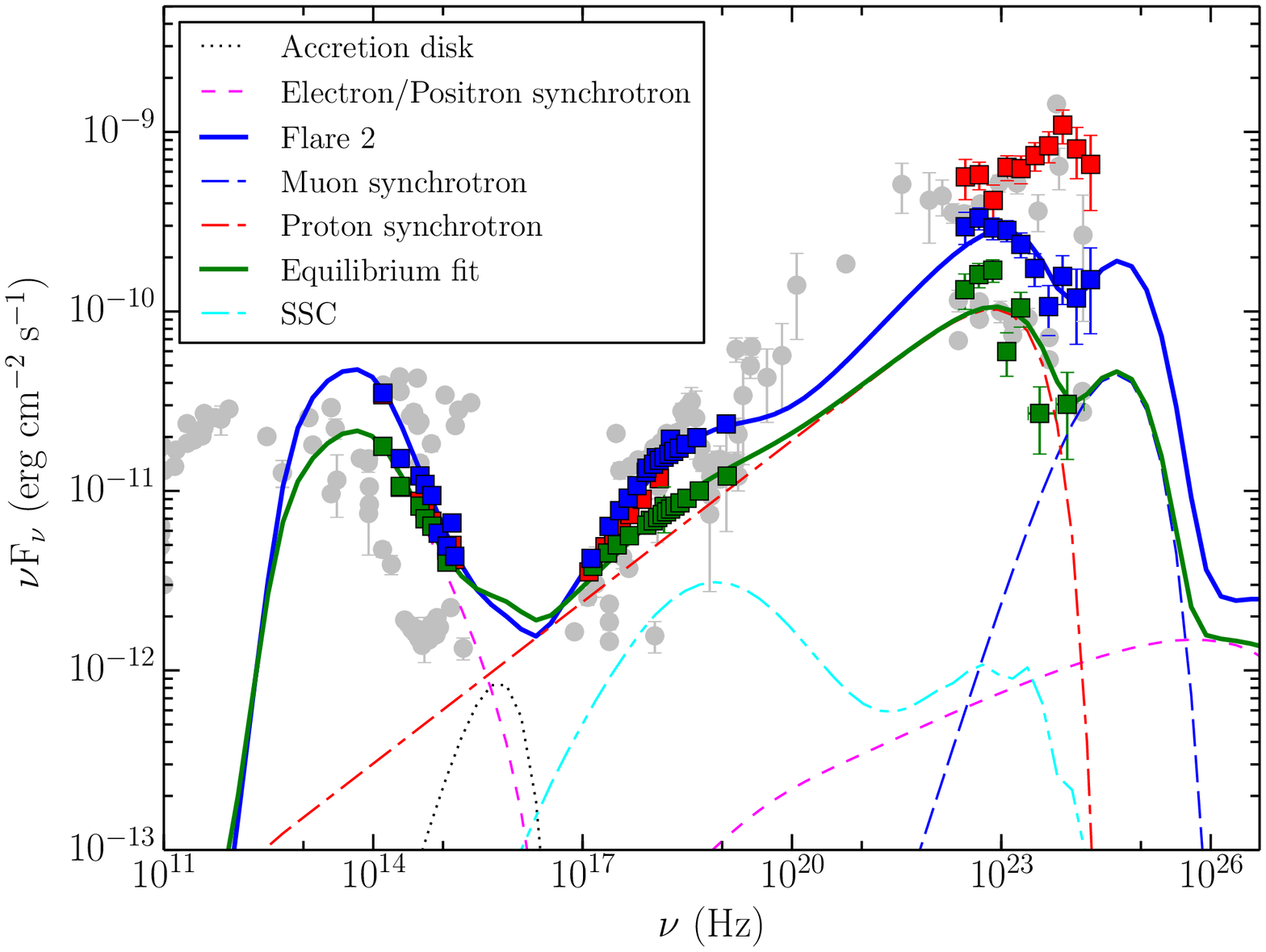}
     }
\caption{Time dependent lepto-hadronic model fits to the multi-wavelength spectral energy distribution of 3C 279. Symbols have their usual meaning as mentioned in Figure~\ref{fig:sed_all}.}\label{fig:hadronic_SED_fit}
\end{figure*}

After the system has reached equilibrium to the SED fitting, we modify one or more of the input parameters as a Gaussian function in time to simulate a flaring scenario as specified in equations 3-5.
The results of the SED generation and modeling, as described above, are presented in Figure~\ref{fig:hadronic_SED_fit} and associated modeling parameters are provided in Table~\ref{tab:parameters} and \ref{tab:flare_parameters}.

\begin{figure*}
\hbox{
      \includegraphics[width=\columnwidth]{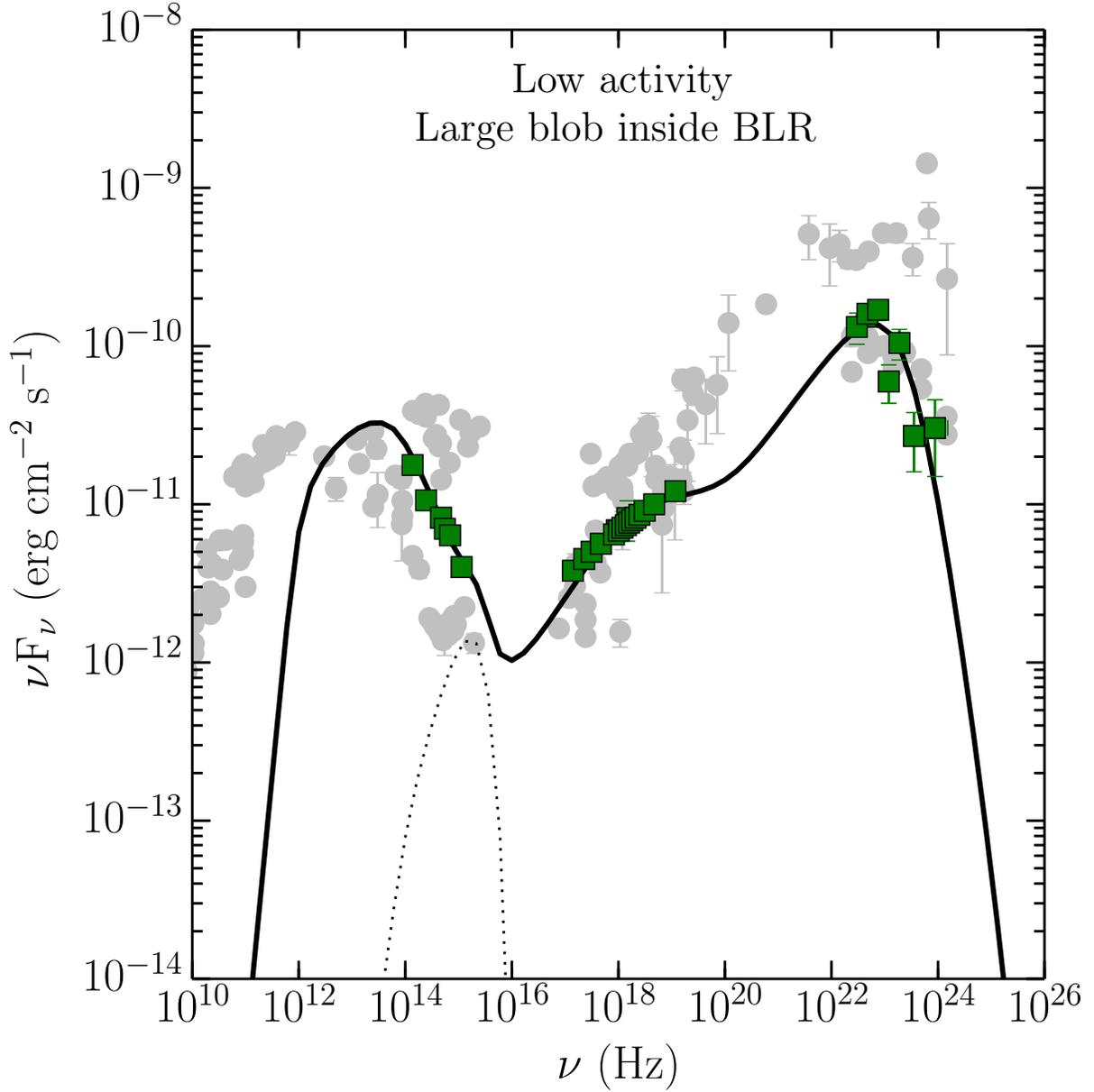}
     }
\caption{Broadband SED of low activity period. Black dotted line represnts the thermal radiation from the accretion disk, whereas solid line is the total radiation predicted by the model. As can be seen, a single-zone leptonic emission model successfully reproduced the observations.}\label{fig:leptonic_SED_fit1}
\end{figure*}

\begin{figure*}\vspace*{-0.7cm}
\hbox{\hspace*{-2cm}
      \includegraphics[width=10.0cm]{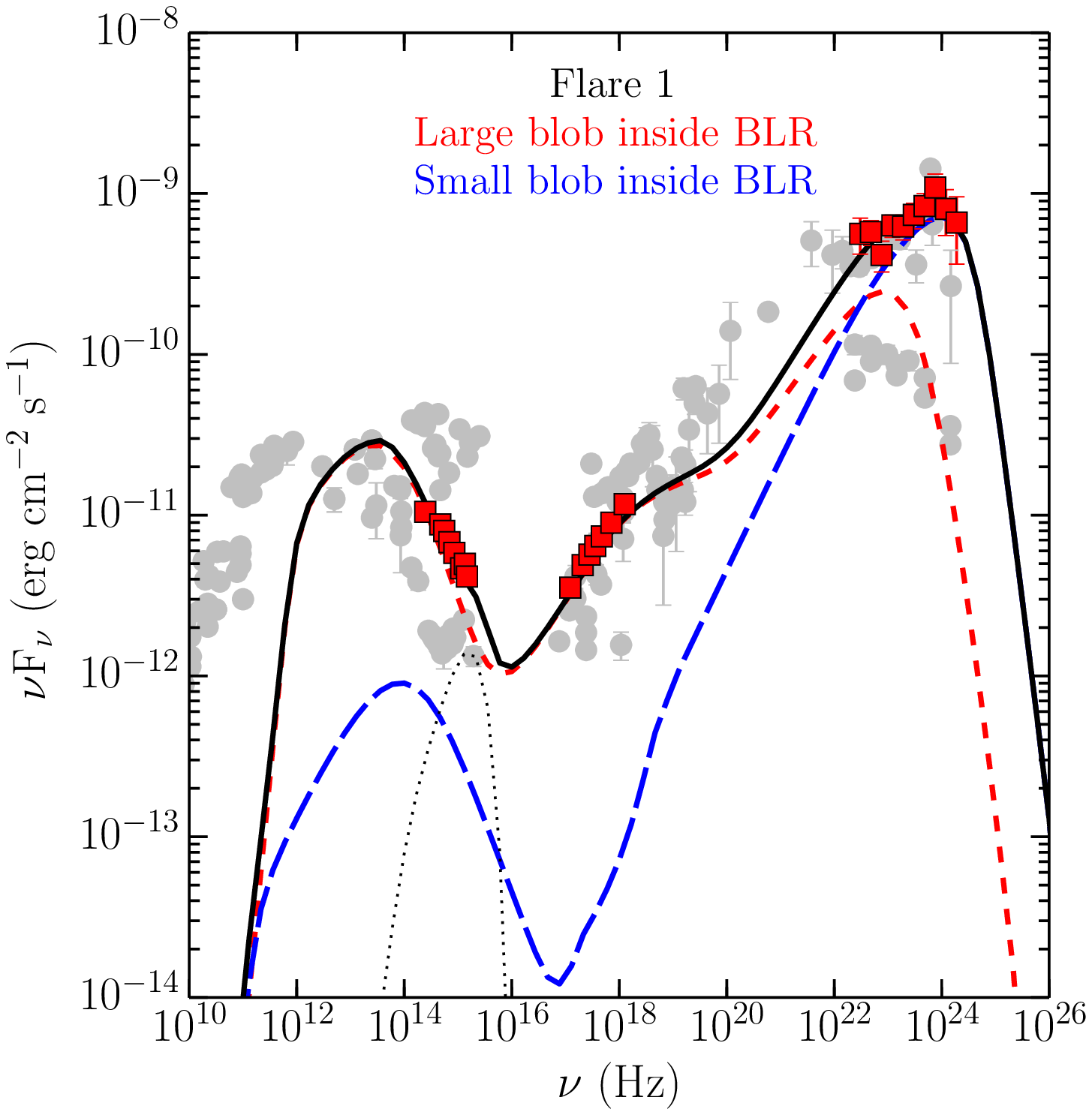}
      \includegraphics[width=10.0cm]{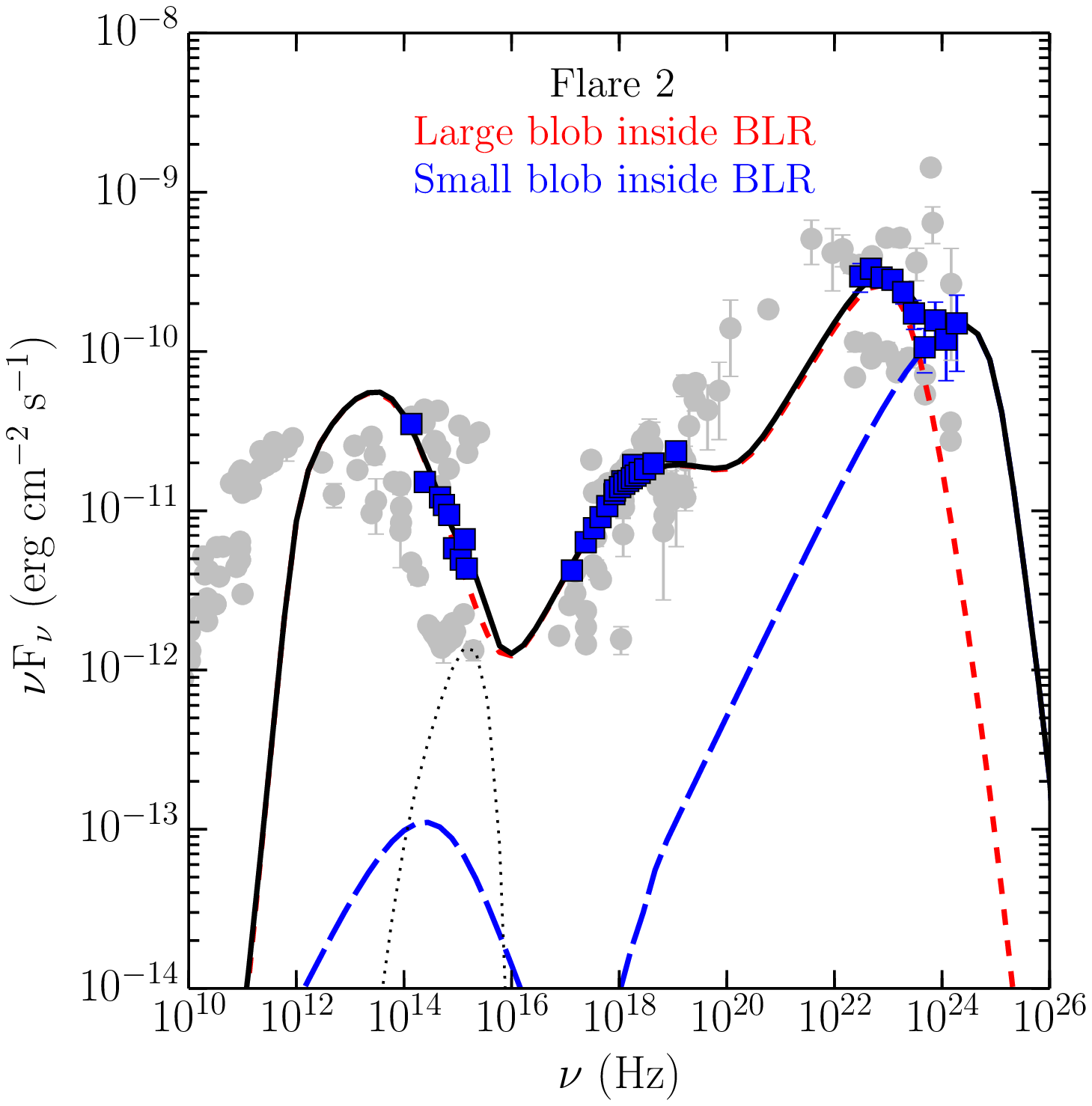}
     }
\hbox{\hspace*{-2cm}
      \includegraphics[width=10.0cm]{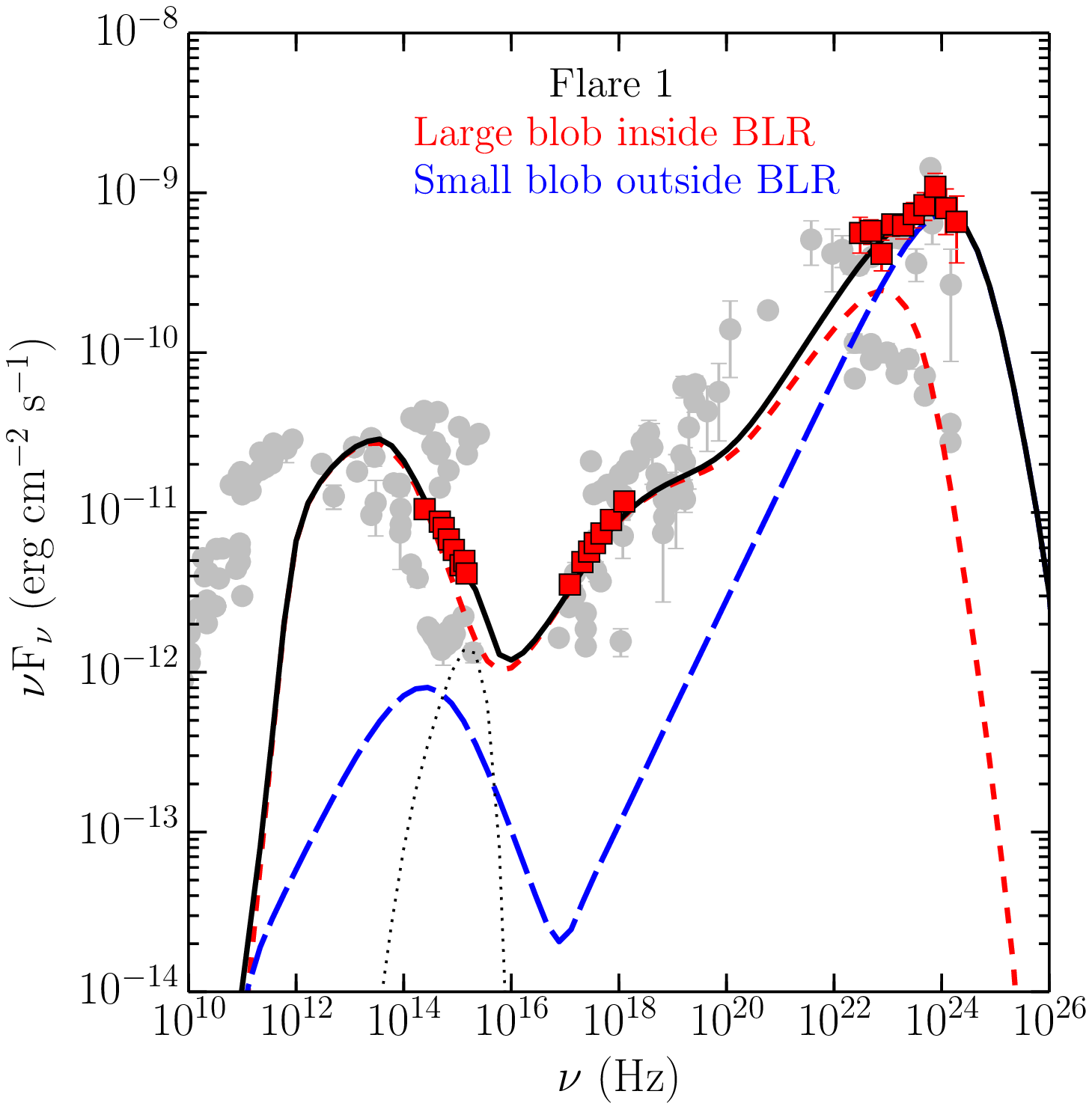}
      \includegraphics[width=10.0cm]{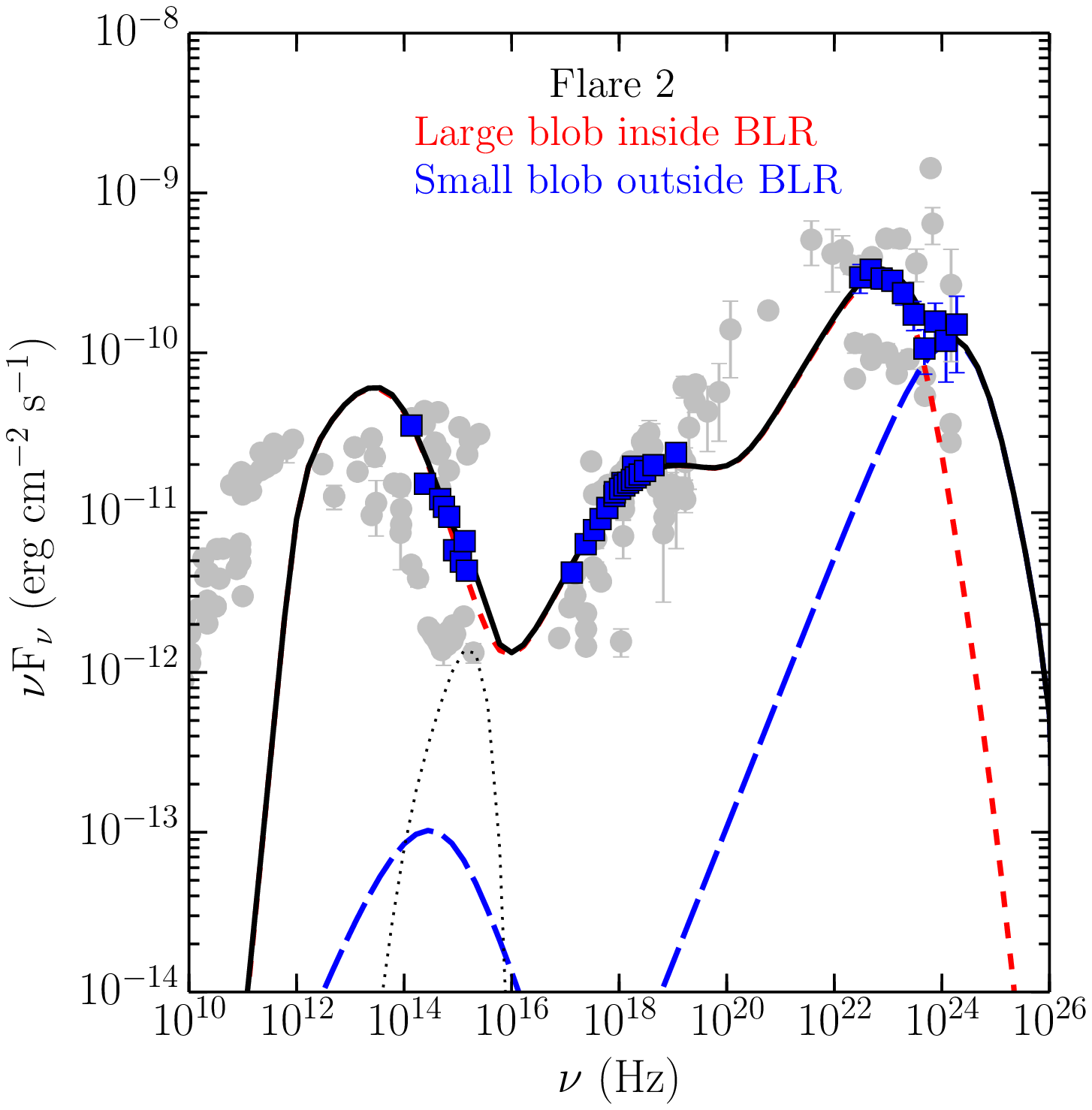}
     }
\caption{Modeled SEDs covering Flare 1 (left column), and Flare 2 (right column). Symbols have the meaning same as in Figure~\ref{fig:sed_all}. Red dashed line represents the total radiation from the large emission region, whereas blue long dahsed line corresponds to that from the small emission region. Solid black line is the sum of all the radiations.}\label{fig:leptonic_SED_fit2}
\end{figure*}

In the leptonic emission scenario, we start with modeling the low activity state where the broadband SED can be reproduced by a single zone approximation. This constrains the physical parameters associated with the large emission region. Further, to explain both Flare 1 and Flare 2 SEDs, we adopt two-zone modeling approach in the light of the above mentioned discussion. The resultant SEDs along with the models are shown in Figure~\ref{fig:leptonic_SED_fit1} and \ref{fig:leptonic_SED_fit2}. The parameters associated with the modeling are provided in Table~\ref{tab:leptonic_flare_parameters}.

We have considered the radiations from two independent regions, and therefore we have more freedom in choosing parameter values since their number is relatively large. However, choice of the parameters is not completely arbitrary and is driven by following constraints. 

{\it Large emission region}$-$ The observed optical-UV SED is primarily emitted by the large region and this not only constrains the shape of the electron energy distribution but also the peak of the synchrotron emission, which in turn controls the location of the IC peak. The properties of the IC radiation are further constrained by the X-ray and $\gamma$-ray spectra. Since we assume that the contribution of the large region to the observed MeV-GeV flux is significantly lower than that by the small emission region, together with its significant contribution to the observed X-ray spectrum, we are able to constrain both the Doppler factor and the magnetic field. Moreover, the observed optical polarization is also assumed to be originated from this region. In this work, we assume the large emission region to be located inside BLR, which is in trend with that generally found for FSRQs \citep{2010MNRAS.402..497G}.

{\it Small emission region}$-$ It is assumed to dominate the observed $\gamma$-ray spectrum but contribute negligibly at optical-UV and X-rays. Adopting the condition that the SSC component should lie below the observed X-ray flux, we are able to constrain the magnetic field (0.1$-$1 G) and Doppler factor ($\delta \gtrsim$ 50). Finally, the electron energy density (and magnetic field also) is determined by the observed $\gamma$-ray flux and by applying the condition that the synchrotron emission lies below the detected optical-UV flux. Lack of VHE observations precludes us to constrain the location of the emission region. As can be seen in Figure~\ref{fig:leptonic_SED_fit2}, the $\gamma$-ray spectrum decreases sharply for the case of inside BLR scenario, due to Klein-Nishina effect. However, if lying outside the BLR, the model predicts significant VHE radiation. For the rest of the physical parameters, associated with both large and small emission regions, we adopt the values typically inferred in FSRQs \citep[e.g.,][]{2010MNRAS.402..497G}.

\section{Discussion}\label{sec:dscsn}
The highest $\gamma$-ray flux during the peak of the flare (Flare 1) is obtained as $\approx$1.2 $\times$ 10$^{-5}$ \phflux, which is similar to that seen during the 2014 April $\gamma$-ray outburst from the same object \citep[e.g.,][]{2015ApJ...803...15P} and comparable to the highest fluxes observed by EGRET instrument on the {\it CGRO} \citep{1998ApJ...497..178W}. The shortest flux doubling timescales are also similar during both epochs. There are, however, a few interesting differences. The $\gamma$-ray spectrum during the 2014 April flare showed a significant curvature \citep{2015ApJ...803...15P}, whereas the 2013 December flare exhibited an extremely hard $\gamma$-ray spectrum. Though, there were no simultaneous X-ray measurements at the peak of the 2013 December flare, the closest \swift-XRT observation reveals the flux level to be similar to that during 2014 April flare. In this aspect, observation of a bright X-ray flare during the Flare 2 phase without a clear $\gamma$-ray counterpart is worth noticing. In the optical band also, 3C 279 does not shows a simultaneous optical and $\gamma$-ray flare. The optical flux appears to peak during the decaying part of the $\gamma$-ray flare (see Figure~\ref{fig:mw_lc}). These observations are in contrast to what was seen during 2014 April flare where flux enhancement was detected across the electromagnetic spectrum \citep{2015ApJ...803...15P}. 

\begin{figure*}
\hbox{
      \includegraphics[width=\columnwidth]{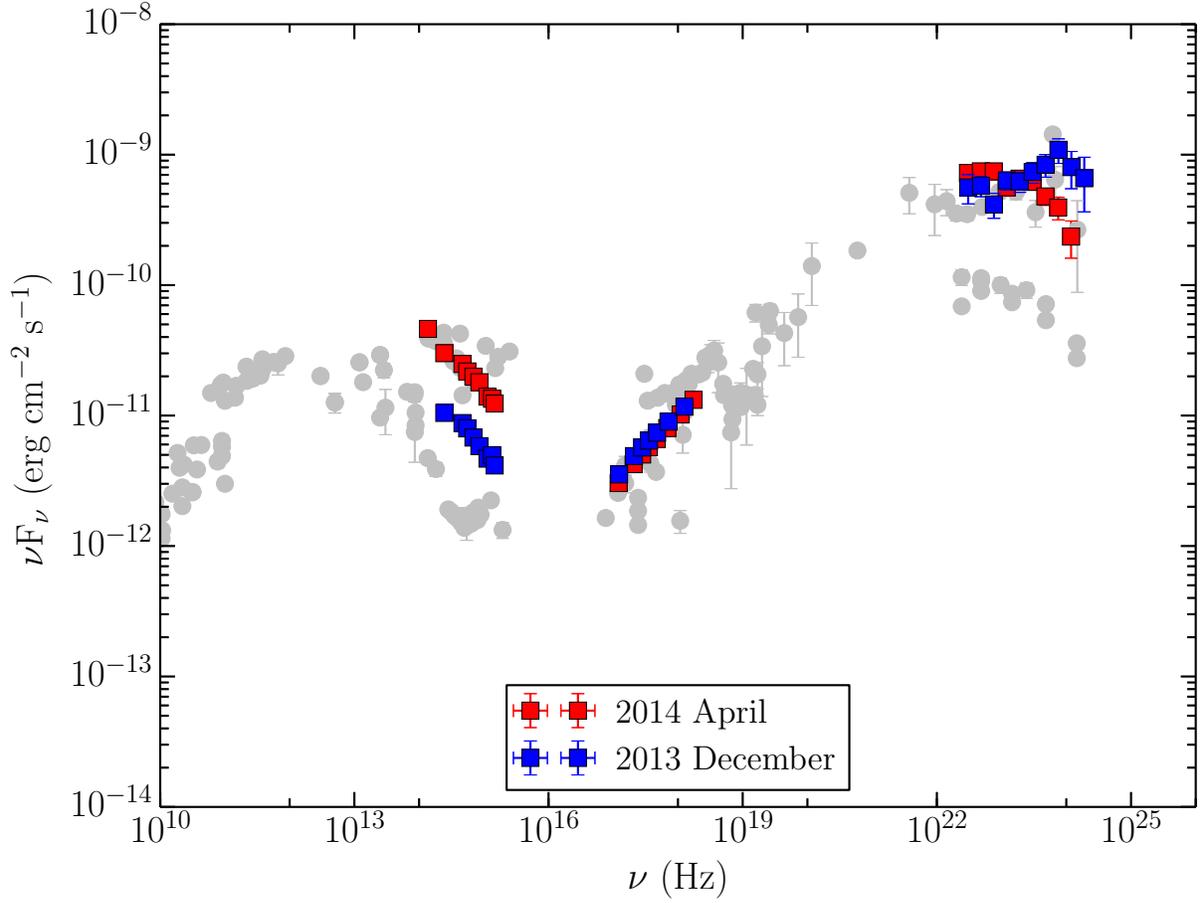}
     }
\caption{Comparison of the broadband SEDs of 3C 279 at the peak of $\gamma$-ray activity during the flares in 2013 December and 2014 April.}\label{fig:SED_Apr_Dec}
\end{figure*}

In Figure~\ref{fig:SED_Apr_Dec}, we compare the SEDs of 3C 279 covering the period of peak $\gamma$-ray activity during 2013 December and 2014 April flares. As can be seen in this plot, flux levels at X-ray and $\gamma$-ray energy bands are similar, whereas, optical flux was lower during the 2013 December event. The shape of optical and X-ray SEDs remains the same during both flares, but what is more interesting is the change in the shape of the $\gamma$-ray spectrum. In the one zone leptonic emission scenario, the shape of the synchrotron spectrum constrains the shape of the high energy $\gamma$-ray radiation, assuming the same population of electrons are responsible for both spectra. If this is the case, a falling optical spectrum should correspond to a steep $\gamma$-ray spectrum, which is observed in 2014 April flare. However, a rising $\gamma$-ray spectrum, as seen during 2013 December flare, is difficult to explain in the light of above mentioned theory, since the optical spectrum is declining. This observation hints that a single zone leptonic emission model may not be able to explain the hard $\gamma$-ray flare detected in 2013 December. Moreover, for an individual $\gamma$-ray flare to occur with no IR/optical/UV flare, is problematic for one zone leptonic modeling. It would seem to suggests that a multi-zone leptonic model or a lepto-hadronic model would work best. The lepto-hadronic model can work since there are two different particle populations being affected by the conditions for the flare. The hadrons respond first, then the leptons would follow suit, producing flares at different times. However, for the case of the flare of 3C 279 in 2013 December, as discussed below, it requires a second hadronic flare to occur alongside the leptonic flare. This would suggest that the initial $\gamma$-ray flare may be its own distinct flare and has no connection with the second flare that follows after. One zone leptonic models, thus, are unlikely to explain this flaring scenario. Therefore, we reproduce these peculiar observations following two independent approaches, a time dependent lepto-hadronic emission scenario and a two-zone leptonic radiation model.

\begin{figure*}
\hbox{\hspace{-1.8cm}
      \includegraphics[width=10cm]{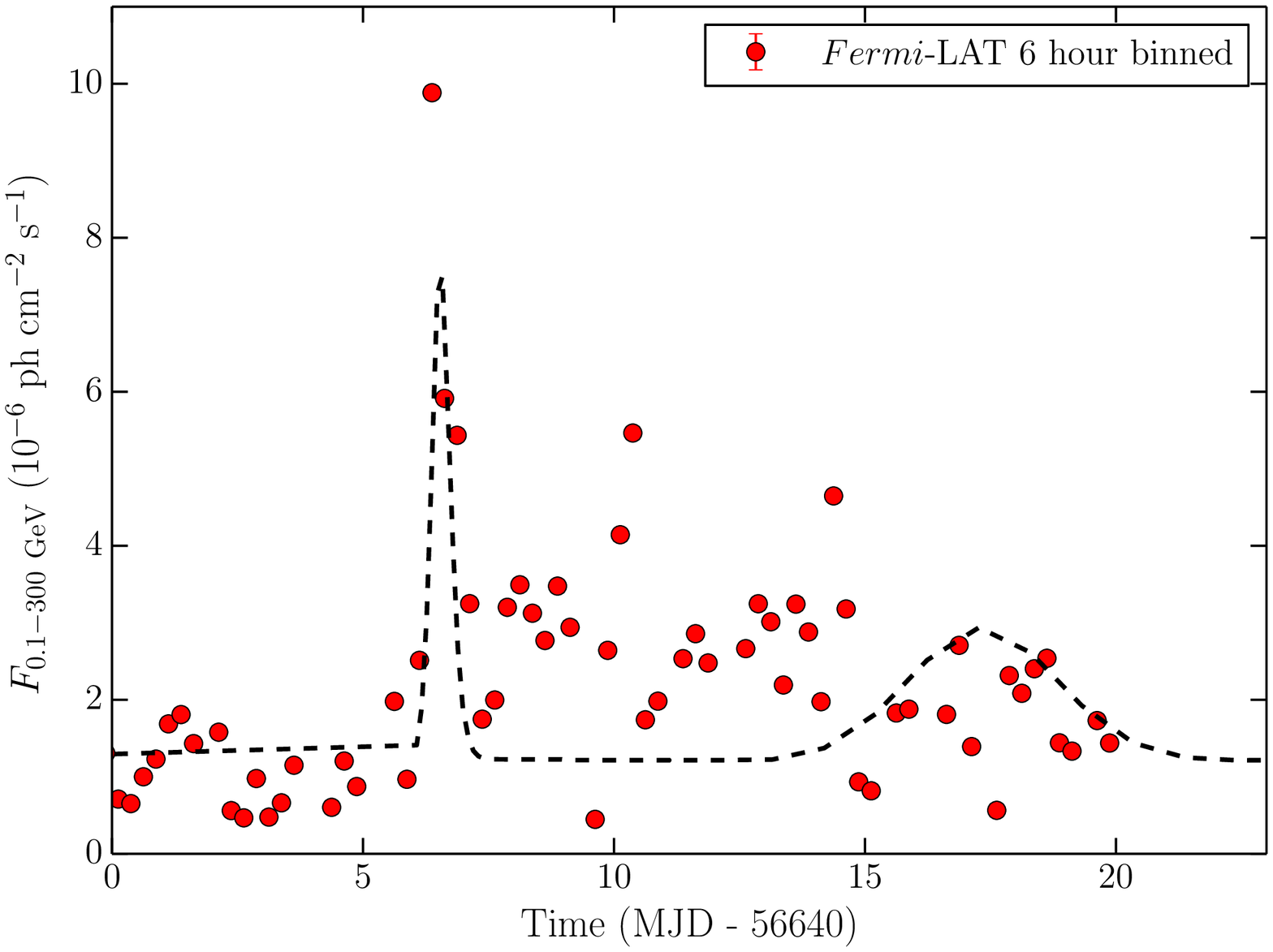}
      \includegraphics[width=10cm]{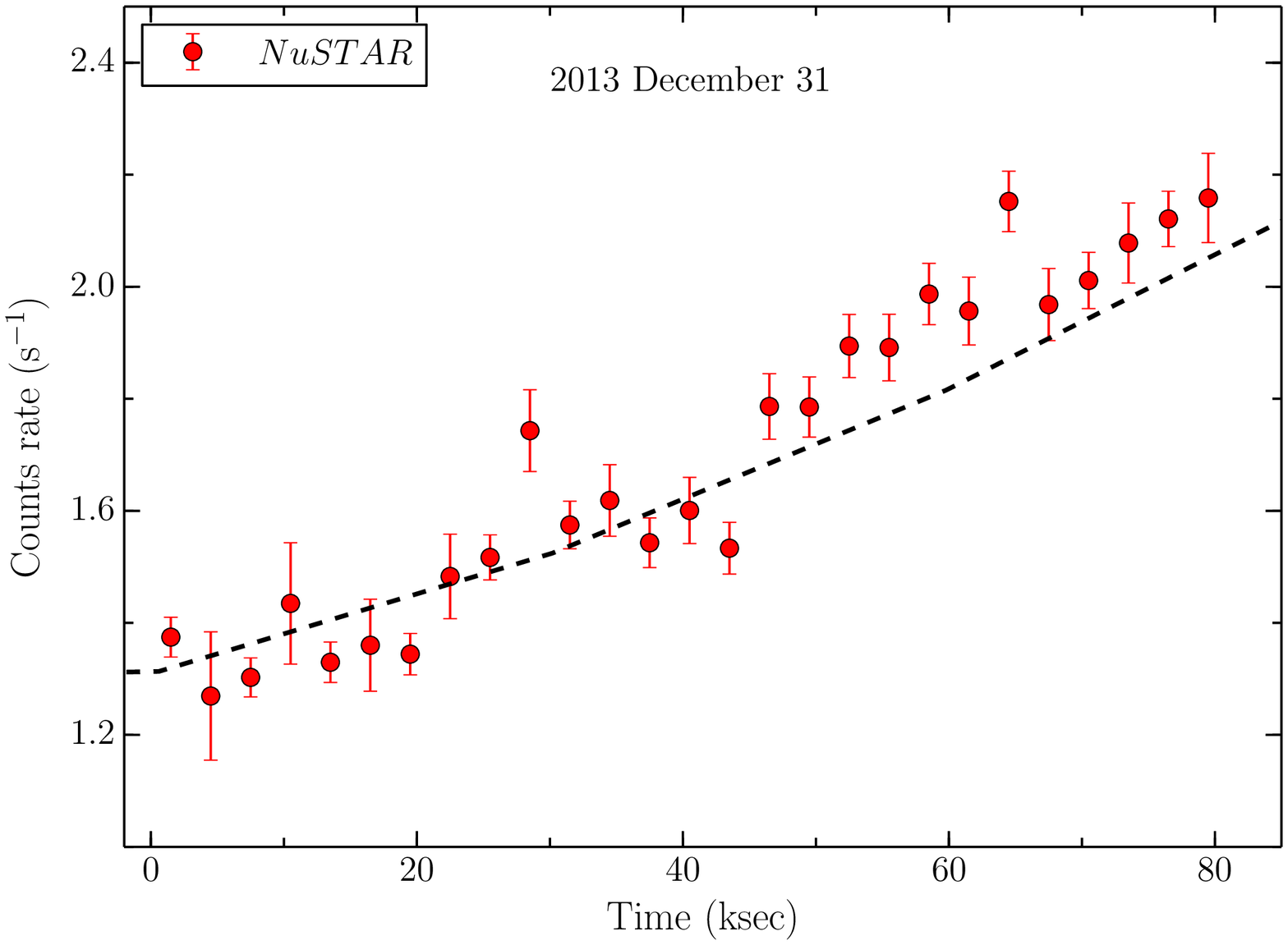}
     }
\hbox{\hspace{3.0cm}
      \includegraphics[width=10cm]{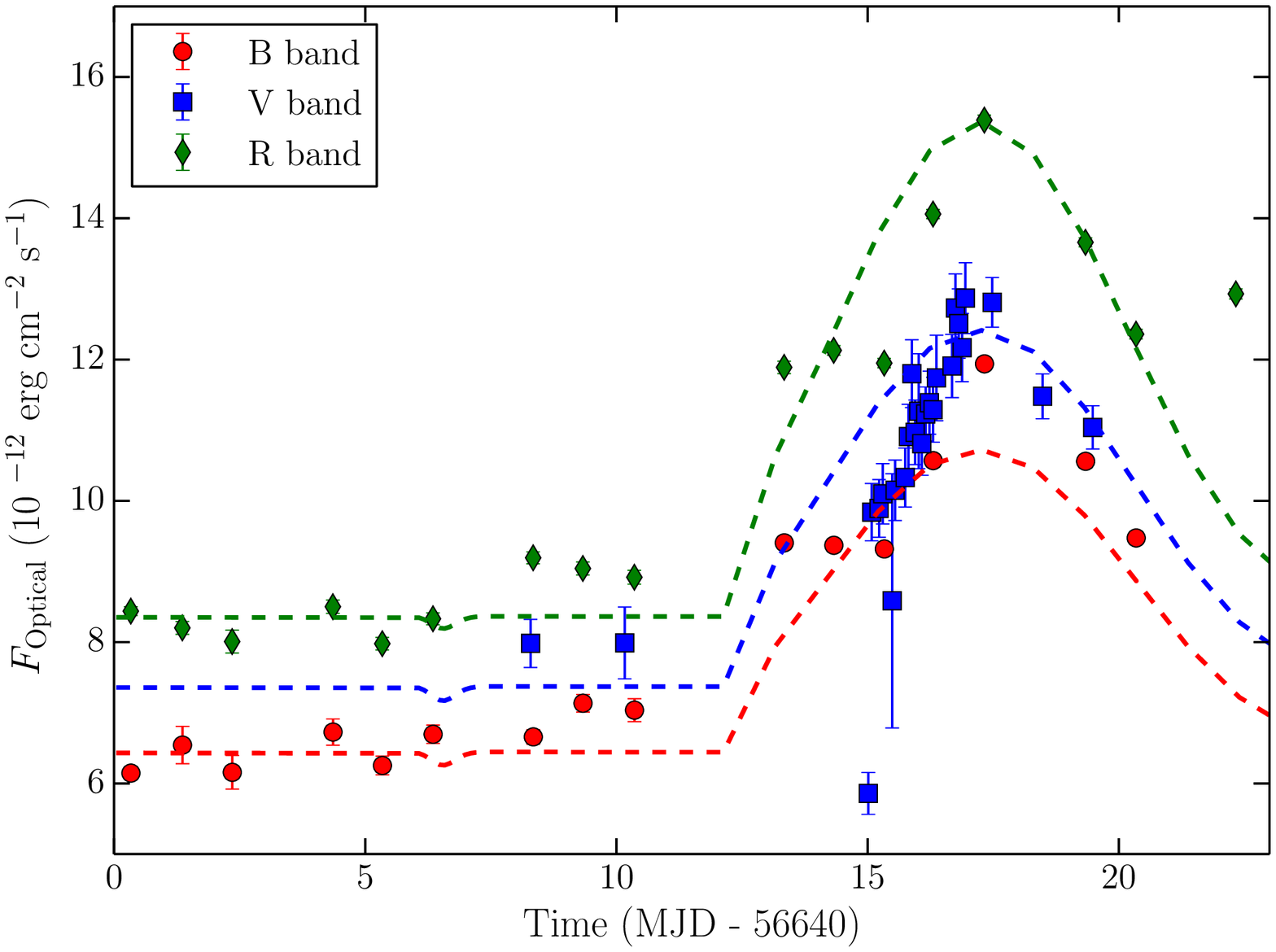}
     }
\caption{Results of the model fitting to the observed multi-wavelength light curves, using the time dependent lepto-hadronic approach described in the text.}\label{fig:lc_modeling}
\end{figure*}

We find that the model fits for the broadband emission for the first flare are satisfactory. When the flare is switched on, with a time step of $\Delta t = 2.0 \times 10^{5}$ s in the comoving frame, there is a gradual increase in both X-ray and $\gamma$-ray emission. The proton spectral index steadily gets harder and conforms with the spectral indices observed in the X-rays in the flaring state. The harder proton spectral index and the increased efficiency of stochastic acceleration pushes the proton synchrotron peak frequency to higher energies. This in turn causes the muon synchrotron emission to increase and produce the extended $\gamma$-ray emission beyond $~10^{24}$ Hz. The $\gamma$-ray flare, centered at MJD 56,646, and the flares in the \swift~B, and V and SMARTS R, centered at MJD 56,656, can also be reproduced by the perturbation setup outlined above. The steady rise in 3$-$79 keV \nustar~light curves for the 2013 December flare can also be explained by our model. However, we obtain poor fits for the 2013 December flare in the 0.3$-$10 keV energy range with our model due to lack of extended coverage during both flaring events. As a result, we omit the figure from this paper (see Figure \ref{fig:lc_modeling}).

After the $\gamma$-ray flare has subsided, the second flare is initiated, with a time step of $\Delta t = 2.0\times10^{6}$ s in the comoving frame. The proton injection spectrum goes through the same perturbation setup and reproduces the weaker $\gamma$-ray flare observed in the Flare 2 state, centered at MJD 56,656. Simultaneously, the electron injection spectrum hardens to produce the increased optical emission. An interesting feature is seen in the spectral break in the X-rays around $\sim 10^{18}$ Hz as measured by \swift-XRT and \nustar during Flare 2. The electron injection luminosity increase produces a marked increase in the emission due to SSC. The combination of increased SSC and proton synchrotron emission reproduces both the increased X-ray emission and the spectral break observed in the Flare 2 state. A counter-intuitive requirement of this scenario is that a softer electron spectral index is needed in order to raise the electron injection luminosity high enough to reproduce adequate fits to the X-ray flux. The data show that in the Flare 2 state the electron synchrotron spectrum is harder compared to the previous flare state. A complete list of flare fitting parameters is given in Table \ref{tab:flare_parameters}.

In summary, we find that the hadronic model presented here produces satisfactory fits for the SED and both the initial $\gamma$-ray flare and the subsequent flaring in the optical and $\gamma$-ray bands following the initial flare. For the X-rays, while we are able to reproduce the \nustar~light curve, we are unable to model the \swift-XRT light curve with the lepto-hadronic model. The lack of coverage and binning in the \swift-XRT band for both flares makes it difficult to reproduce the light curves in this band. A perturbation of the proton spectral index and the efficiency of stochastic acceleration explain the initial hard $\gamma$-ray flare. An increase in the injection luminosity and a change to the electron injection spectral index produces adequate fits to the optical and X-ray spectra and light curves. The combination of electron SSC and proton synchrotron explains the spectral break observed in 3C 279 in Flare 2 state following the initial 2013 December flare. A perturbation to the proton injection similar to the initial $\gamma$-ray flare then explains the flaring in the $\gamma$-ray band during Flare 2. In the context of hadronic modeling, proton synchrotron explains the emission from soft X-rays to GeV $\gamma$-rays. By lowering the magnetic fields and increasing the proton and electron injection luminosities, substantial SSC emission can take place, centered around soft to hard X-rays. The combination of proton synchrotron and SSC emission can explain the increased X-ray flux and the spectral break observed with \swift-XRT and \nustar. This scenario requires that the jet becomes very strongly particle dominated. The magnetic field is $\sim30$ G, and the ratio of particle kinetic energy to magnetic energy is $\epsilon_{pB} \sim 2.0 \times 10^{3}$. Given estimated location of the emission region, $R_{axis} = 0.019~pc \approx 5.84 \times 10^{16}~cm $, the particle dominated scenario conflicts with a Poynting flux dominated scenario expected at these locations close to the supermassive black hole (e.g., \citealt{2012MNRAS.423.3083M}, but see \citealt{2015MNRAS.451..927Z}). However, the synchrotron and SSC spectral components that make up the broadband emission for the one-zone lepto-hadronic model does not depend on the location of the emission region. The observed variability time scale can be caused by other physical processes further down the jet where the emission zone can be particle dominated. Furthermore, as the mechanisms of jet launching, acceleration, and collimation are still poorly understood, we do not believe that a strongly particle-dominated jet can be completely ruled out.  

Both the two-zone leptonic and the lepto-hadronic emission models used in this work are able to produce acceptable fits to the observed SEDs of 3C 279. However, the physical parameters returned by the above two model fits are not in agreement to each other. Future polarimetric observations at X-ray and $\gamma$-ray energies will certainly help in distinguishing between these two radiative processes working in blazar jets \citep[][]{2013ApJ...774...18Z}.

The multi-wavelength observations of 3C 279 during the twin $\gamma$-ray flares in 2013 December and 2014 April have revealed many peculiar features. The spectral characteristics of the source were found to be same in optical-UV and X-rays during both the activity states, however, $\gamma$-ray observations clearly show a change in the spectrum over the course of 4 months. Moreover, the multi-wavelength variability behavior of the source also changed dramatically with an uncorrelated flux variations seen in 2013 December and a simultaneous flux enhancement during 2014 April flare. As discussed above, 2013 December flare cannot be explained by one-zone leptonic models and we need to look for alternative approaches such as multi-zone leptonic or lepto-hadronic modeling. These observations hint for the presence of a variety of radiative processes working at different activity levels in the jet, of which, these twin flares are just an example. Overall, the observations reflect the level of complexity involved in understanding the physical properties of blazars and we do need to carry continuously monitoring using broadband observational facilities as close as possible in time, for better understanding of the these peculiar objects.

\section{Summary}\label{sec:summary}
In this work, we study the giant $\gamma$-ray flare observed from 3C 279 in 2013 December. Our main findings are as follows
\begin{itemize}
\item The highest $\gamma$-ray flux, in the energy range of 0.1$-$300 GeV, is found to be (1.22$\pm$0.25) $\times$ 10$^{-5}$ \phflux. This is similar to the flux level seen during the 2014 April flare, however, less than that observed during 2015 June $\gamma$-ray outburst \citep{2015ApJ...808L..48P}. The associated photon index is extremely hard and has a value of 1.70$\pm$0.13. The fastest $\gamma$-ray flux doubling time of 3.04$\pm$0.77 hr is also detected. These results are inline with the findings of \citet{2015ApJ...807...79H}.
\item During the Flare 2, a bright X-ray flare is observed with flux doubling time as short as 2.89$\pm$0.67 hr. This is probably the first report of such an extremely fast X-ray variability seen from 3C 279.
\item Unlike 2014 April outburst, the $\gamma$-ray flare during 2013 December exhibits a hard $\gamma$-ray spectrum and an uncorrelated multi-wavelength variability behavior. These observations are difficult to explain by the commonly accepted one-zone leptonic emission scenario, and thus alternatives such as time dependent lepto-hadronic and multi-zone leptonic radiative models are proposed to explain the observed phenomena.

\item The observed SEDs, optical U, B, V, \nustar, and \fermi~$\gamma$-ray light curves can be explained by a one-zone lepto-hadronic model. However, we obtain poor fits to the \swift-XRT light curve during both flaring events. Synchrotron emission from a distribution of electrons/positrons and protons produces the IR-UV and intermediate to high energy $\gamma$-rays. A combination of electron synchrotron self Compton and proton synchrotron produced adequate fits to the soft to hard X-rays. Synchrotron emission from secondary particles generated through photo-hadronic interactions between the protons and photons generates a very high energy spectral component beyond $20~GeV$. 

\item For the 2013 December flare, an initial spectral hardening of the proton distribution and an increase in it's acceleration efficiency was used. This produces the initial harder $\gamma$-ray spectrum observed in the SED and the flaring in the \fermi~band. A second spectral hardening for the proton distribution follows, with a spectral softening of the electron distribution to explain the flaring after the initial $\gamma$-ray flare. A combination of the increase in electron SSC and proton synchrotron adequately produces the X-ray spectral break observed in the Flare 2 state.   

\item In the context of a one zone lepto-hadronic model, the hadrons are compressed at a shock front and obtain energy due to Fermi II processes from the increased turbulence downstream. The protons are accelerated much more efficiently than the electrons, which could explain the lack of an optical flare to coincide with the 2013 December $\gamma$-ray flare. Both particle populations are compressed from a shock front and increase in energy and luminosity during the 2014 April flare. This suggests that the two flaring events are likely not correlated.  

\item The observed SEDs can also be successfully reproduced by a two-zone leptonic radiative model. In this approach, a large emission region is found to emit IR to X-rays, whereas $\gamma$-ray emission is explained by a relatively fast moving small emission region. Due to unavailability of VHE observations we could not constrain the location of the $\gamma$-ray emitting region.
\end{itemize} 

\acknowledgments
We are thankful to the referee for providing a constructive report. This research has made use of data, software and/or web tools obtained from NASA’s High Energy Astrophysics Science Archive Research Center (HEASARC), a service of Goddard Space Flight Center and the Smithsonian Astrophysical Observatory. Part of this work is based on archival data, software, or online services provided by the ASI Science Data Center (ASDC). This research has made use of the XRT Data Analysis Software (XRTDAS) developed under the responsibility of the ASDC, Italy. This research has also made use of the NuSTAR Data Analysis Software (NuSTARDAS) jointly developed by the ASI Science Data Center (ASDC, Italy) and the California Institute of Technology (Caltech, USA). Data from the Steward Observatory spectropolarimetric monitoring project were used. This program is supported by Fermi Guest Investigator grants NNX08AW56G, NNX09AU10G, and NNX12AO93G. This paper has made use of up-to-date SMARTS optical/near-infrared light curves that are available at www.astro.yale.edu/smarts/glast/home.php. Use of {\it Hydra} cluster at the Indian Institute of Astrophysics is acknowledged. The work of M.B. is supported by the South African Research Chair Initiative (SARChI) of the Department of Science and Technology and the National Research Foundation\footnote{Any opinion, finding and conclusion or recommendation expressed in this material is that of the authors and the NRF does not accept any liability in this regard.} of South Africa.

\bibliographystyle{apj}
\bibliography{Master}

\newpage
\begin{table}
\caption{Results of the $\gamma$-ray spectral fitting of 3C 279, obtained for different activity states. Col.[1]: period of observation (MJD); Col.[2]: activity state; Col.[3]: model used (PL: power law, LP: logParabola); Col.[4]: integrated $\gamma$-ray flux (0.1$-$300 GeV), in units of 10$^{-6}$ \phflux; Col.[5] and [6]: spectral parameters; Col.[7]: test statistic; Col.[8]: $TS_{\rm curve}$ .}\label{tab:gamma_spec}
\begin{center}
\begin{tabular}{cccccccc}
\hline
Period & Activity & Model & $F_{0.1-300~{\rm GeV}}$ & $\Gamma_{0.1-300~{\rm GeV}}/\alpha$ & $\beta$ & TS & $TS_{\rm curve}$\\
~[1] & [2] & [3] & [4] & [5] & [6] & [7] & [8] \\
\hline
56,640$-$56,645 & Low activity & PL & 0.84 $\pm$ 0.08 & 2.48 $\pm$ 0.09 &                 & 403.06  & \\
                &              & LP & 0.81 $\pm$ 0.08 & 2.38 $\pm$ 0.12 & 0.09 $\pm$ 0.08 & 403.13  & 1.47 \\
56,645$-$56,649 & Flare 1      & PL & 3.76 $\pm$ 0.30 & 1.95 $\pm$ 0.05 &                 & 1245.89 & \\
                &              & LP & 3.30 $\pm$ 0.33 & 1.70 $\pm$ 0.12 & 0.09 $\pm$ 0.04 & 1272.25 & 6.46 \\
56,649$-$56,660 & Flare 2      & PL & 1.90 $\pm$ 0.14 & 2.28 $\pm$ 0.06 &                 & 991.19  & \\
                &              & LP & 1.79 $\pm$ 0.15 & 2.15 $\pm$ 0.10 & 0.07 $\pm$ 0.04 & 994.27  & 3.20 \\
\hline
\end{tabular}
\end{center}
\end{table}

\begin{table*}
\begin{center}
{
\scriptsize
\caption{Results of the analysis performed to generate broadband SEDs. \fermi-LAT analysis results are presented in Table~\ref{tab:gamma_spec}.}\label{tab:sed_flux}
\begin{tabular}{ccccccc}
\tableline
 & & & \nustar  & & &\\
 Activity state & Exp.\tablenotemark{a} & $\Gamma_{3-79~{\rm keV}}$\tablenotemark{b} & $F_{3-79~{\rm keV}}$\tablenotemark{c} & Norm.\tablenotemark{d} & Stat.\tablenotemark{e} & \\
 \tableline
 Low activity & 40 & 1.74$^{+0.03}_{-0.03}$ & 3.20$^{+0.08}_{-0.10}$ & 2.94$^{+0.17}_{-0.16}$ & 509.62/470 & \\
 Flare 2      & 43 & 1.76$^{+0.02}_{-0.02}$ & 2.31$^{+0.03}_{-0.03}$ & 6.21$^{+0.24}_{-0.23}$ & 658.70/688 & \\
\tableline
 & & & \swift-XRT  & & &\\
 Activity state & Exp.\tablenotemark{a} & $\Gamma_{0.3-10~{\rm keV}}$\tablenotemark{f} & $F_{0.3-10~{\rm keV}}$\tablenotemark{g} & Norm.\tablenotemark{d} & Stat.\tablenotemark{e} & \\
 \tableline
 Low activity & 2.14 & 1.68$^{+0.11}_{-0.11}$ & 2.04$^{+0.20}_{-0.18}$ & 2.89$^{+0.23}_{-0.23}$ & 19.98/27 & \\
 Flare 1      & 3.89 & 1.50$^{+0.06}_{-0.06}$ & 2.78$^{+0.17}_{-0.17}$ & 3.32$^{+0.16}_{-0.16}$ & 59.06/71 & \\
 Flare 2      & 28.10& 1.44$^{+0.02}_{-0.02}$ & 3.61$^{+0.07}_{-0.07}$ & 4.05$^{+0.07}_{-0.07}$ & 345.10/355& \\
 \tableline
 & & & {\it Swift}-UVOT &  & & \\
 Activity state & V\tablenotemark{h} & B\tablenotemark{h} & U\tablenotemark{h} & UVW1\tablenotemark{h} & UVM2\tablenotemark{h} & UVW2\tablenotemark{h} \\
 \tableline
Low activity  & -- & -- & -- & 4.01 $\pm$ 0.22 & -- & -- \\
Flare 1 & 7.98 $\pm$ 0.34&6.78 $\pm$ 0.26&5.83 $\pm$ 0.24&4.68 $\pm$ 0.26&4.98 $\pm$ 0.26&4.15 $\pm$ 0.23 \\
Flare 2 &10.87 $\pm$ 0.12& --            &5.81 $\pm$ 0.30&4.93 $\pm$ 0.30&6.63 $\pm$ 0.09&4.33 $\pm$ 0.24 \\
\tableline
 & & & SMARTS &  & & \\
 Activity state & B\tablenotemark{i} & V\tablenotemark{i} & R\tablenotemark{i} & J\tablenotemark{i} & K\tablenotemark{i} &  \\
 \tableline
Low activity  & 6.37 $\pm$ 0.09 & 7.00 $\pm$ 0.06 & 8.23 $\pm$ 0.05 & 10.60 $\pm$ 0.05 & 17.71 $\pm$ 0.05 &  \\
Flare 1 & -- & -- & 8.76 $\pm$ 0.06 & 10.48 $\pm$ 0.07 & 34.31 $\pm$ 0.15 &  \\
Flare 2 & 9.42 $\pm$ 0.03 & -- & 12.13 $\pm$ 0.03 & 15.16 $\pm$ 0.04 & 35.05 $\pm$ 0.06 &  \\
 \tableline
\end{tabular}
\tablenotetext{a}{Net exposure in kiloseconds}
\tablenotetext{b}{Photon index of the power law model in 3$-$79 keV energy range.}
\tablenotetext{c}{Power law flux in 3$-$79 keV energy range, in units of 10$^{-11}$ \ergflux.}
\tablenotetext{d}{Normalization at 1 keV in 10$^{-3}$ \phflux~keV$^{-1}$.}
\tablenotetext{e}{Statistical parameters: $\chi^2$/dof.}
\tablenotetext{f}{Photon index of the absorbed power law model in 0.3$-$10 keV energy range.}
\tablenotetext{g}{Unabsorbed flux in units of 10$^{-11}$ erg cm$^{-2}$ s$^{-1}$, in 0.3$-$10 keV energy band.}
\tablenotetext{h}{Average flux in {\it Swift} V, B, U, W1, M2, and W2 bands, in units of 10$^{-12}$ erg cm$^{-2}$ s$^{-1}$.}
\tablenotetext{i}{Average flux in SMARTS B, V, R, J, and K bands, in units of 10$^{-12}$ erg cm$^{-2}$ s$^{-1}$.}
}
\end{center}
\end{table*}

\begin{table*}
\begin{center}
\caption[]{Lepto-hadronic parameter values used for the equilibrium fit to the SEDs of 3C279}\label{tab:parameters}

\begin{tabular}{lcc}
\hline
Parameter & Symbol & Value \\
\hline
 Magnetic field  & $ B $ & $30$ G \\ 
 Radius of emission region & $ R $ & $1.62 \times 10^{15}$~cm \\
 Constant multiple for escape timescale & $\eta$ & 11.0 \\
 Bulk Lorentz factor & $\Gamma$ & 18 \\
 Observing angle & $\theta_{\rm obs}$ & $5.5 \times 10^{-2}$~rad \\
 Minimum proton Lorentz factor & $ \gamma_{\rm p, min} $ & $ 1.0 $ \\
 Maximum proton Lorentz factor & $ \gamma_{\rm p, max} $ & $ 4.5 \times 10^{8} $ \\
 Proton injection spectral index & $q_{p}$ & 2.4 \\
 Proton injection luminosity & $L_{\rm p, inj}$ & $9.84 \times 10^{46}$~erg~s$^{-1}$ \\
 Minimum electron Lorentz factor & $ \gamma_{\rm e, min} $ & $ 1.72 \times 10^{2} $ \\
 Maximum electron Lorentz factor & $ \gamma_{\rm e, max} $ & $ 2.0 \times 10^{3} $ \\
 Electron injection spectral index & $q_{e}$ & 3.6 \\
 Electron injection luminosity & $L_{\rm e, inj}$ & $4.41 \times 10^{41}$~erg~s$^{-1}$ \\
 Supermassive black hole mass & $M_{\rm BH}$ & $3.0 \times 10^{8} \, M_{\odot}$ \\
 Eddington ratio & $l_{\rm Edd}$ & $5.15 \times 10^{-2}$ \\
 Blob location along the jet axis & $R_{\rm axis}$ & 0.019~pc \\
 Ratio between the acceleration and escape timescales & $ t_{\rm acc}/t_{\rm esc} $ & $5.0$ \\
\hline
\end{tabular}
\end{center}
\end{table*}

\begin{table*}
\centering
\caption{Model light curve fit parameters obtained from the lepto-hadronic modeling. The negative value for the electron spectral index to model the second flare represents a hardening of the spectral index. The negative value for the electron injection luminosity indicates an increase in electrons to model the flaring state.}
\label{tab:flare_parameters}
\begin{tabular}{lcccc}
\hline
 Scenario & $\sigma [s]$ & $K_{q}$ & $K_{acc}$ & $K_{L}$ \\
\hline
Proton (1st Flare) & $4.0 \times 10^{5}$ & $-0.3$ & $3.0$ & $0.95$ \\
Proton (2nd Flare) & $4.0 \times 10^{6}$ & $-0.25$ & $1.0$ & $2.8$ \\
Electron (2nd Flare) & $4.0 \times 10^{6}$ & $0.35$ & $4.5$ & $-0.57$ \\
\hline
\end{tabular} \\ 
\end{table*}

\begin{table*}
\caption{The SED model parameters obtained by adopting two-zone leptonic emission scenario. The black hole mass and the accretions disk luminosity are taken as 3 $\times$ 10$^{8}$ $M_{\odot}$ and 2 $\times$ 10$^{45}$ \lum, respectively. Viewing angle and characteristic temperature of the IR torus are taken as 1$^{\circ}$ and 1000 K, respectively. The minimum Lorentz factor of the emitting electrons is adopted as unity.}
\label{tab:leptonic_flare_parameters}
\centering\scriptsize
\begin{tabular}{llcccccccccccccc}
\hline
Activity State & Location & $R_{\rm size}$ & $\delta$ & $\Gamma$ & $B$ & $\gamma_{\rm brk}$ & $\gamma_{\rm max}$ & $p$ & $q$ & $U_{\rm e}$ & $P_{\rm e}$ & $P_{\rm p}$ & $P_{\rm B}$ & $P_{\rm r}$\\
\hline
Low & Large region inside BLR & 1 $\times$ 10$^{16}$ & 17 & 9 & 2.9 & 579 & 3 $\times$ 10$^{4}$ & 1.9 & 5.2 & 0.27 & 44.26 & 46.61 & 44.35 & 44.87\\
\hline
Flare 1 & Large region inside BLR & 1 $\times$ 10$^{16}$ & 18 & 9 & 2.0 & 685 & 3 $\times$ 10$^{4}$ & 1.9 & 5.2 & 0.36 & 44.41 & 46.76 & 44.06 & 44.97\\
~& Small region inside BLR & 5 $\times$ 10$^{15}$ & 54 & 40 & 0.55 & 1139 & 5 $\times$ 10$^{4}$ & 1.6 & 4.8 & 0.003 & 43.15 & 45.02 & 43.77 & 44.84\\
~& Small region outside BLR & 5 $\times$ 10$^{15}$ & 56 & 45 & 0.20 & 3140 & 5 $\times$ 10$^{4}$ & 1.6 & 4.8 & 0.006 & 43.57 & 45.26 & 43.02 & 44.97\\
\hline
Flare 2 & Large region inside BLR & 1 $\times$ 10$^{16}$ & 18 & 9 & 2.9 & 514 & 3 $\times$ 10$^{4}$ & 1.6 & 5.2 & 0.21 & 44.22 & 46.24 & 44.42 & 45.05\\
~& Small region inside BLR & 5 $\times$ 10$^{15}$ & 47 & 30 & 0.25 & 3053 & 5 $\times$ 10$^{4}$ & 1.6 & 4.8 & 0.002 & 42.54 & 44.23 & 42.72 & 44.00\\
~& Small region outside BLR & 5 $\times$ 10$^{15}$ & 47 & 30 & 0.12 & 4407 & 5 $\times$ 10$^{4}$ & 1.6 & 4.8 & 0.005 & 42.98 & 44.61 & 42.08 & 43.90\\
\hline
\end{tabular}
\tablecomments{The parameters are as follows. $R_{\rm size}$: size of the emission region in cm, $\delta$ is the Doppler factor, $\Gamma$: bulk Lorentz factor, $B$: magnetic field (Gauss), $\gamma_{\rm brk}$ and $\gamma_{\rm max}$: the break and maximum Lorentz factors, $p$ and $q$: the slopes of the underlying broken power law electron energy distribution, $U_{\rm e}$: particle energy density (erg cm$^{-3}$). The last four columns report the jet power in electrons, protons (assumed to be cold and having equal electron number density), magnetic field, and in radiation, in logarithmic scale.}
\end{table*}

\newpage

\newpage

\newpage

\newpage

\newpage

\newpage

\newpage

\newpage

\newpage

\newpage

\end{document}